\documentclass[aip,jcp,amsmath,amssymb,floatfix,citeautoscript,reprint]{revtex4-1}

\usepackage{graphicx,graphics,epsfig,subfigure,times,bm,bbm,amssymb,amsmath,amsfonts,amsthm,mathrsfs,MnSymbol}
\usepackage[matrix,frame,arrow]{xypic}
\usepackage[normalem]{ulem}
\usepackage{slashed}
\usepackage{dcolumn}
\usepackage{tabularx}
\usepackage{color}
\usepackage[usenames,dvipsnames,svgnames,table]{xcolor}
\usepackage[english]{babel}
\usepackage{verbatim}
\definecolor{darkblue}{rgb}{0.0,0.0,0.3}
\usepackage[colorlinks=true,
            linkcolor=red,
            urlcolor= darkblue,
            citecolor=blue]{hyperref}

\newcommand{\bea}{\begin{eqnarray}}
\newcommand{\eea}{\end{eqnarray}}

\begin{document}

\title{2D spectroscopies from condensed phase dynamics: Accessing third-order response properties from equilibrium multi-time correlation functions}

\author{Kenneth A. Jung}
\author{Thomas E. Markland}
\email{tmarkland@stanford.edu}
\affiliation{Department of Chemistry, Stanford University, Stanford, California, 94305, USA}

\date{\today}

\begin{abstract}
The third-order response lies at the heart of simulating and interpreting nonlinear spectroscopies ranging from two dimensional infrared (2D-IR) to 2D electronic (2D-ES), and 2D sum frequency generation (2D-SFG). The extra time and frequency dimensions in these spectroscopies provides access to rich information on the electronic and vibrational states present, the coupling between them, and the resulting rates at which they exchange energy that are obscured in linear spectroscopy, particularly for condensed phase systems that usually contain many overlapping features. While the exact quantum expression for the third-order response is well established it is incompatible with the methods that are practical for calculating the atomistic dynamics of large condensed phase systems. These methods, which include both classical mechanics and quantum dynamics methods that retain quantum statistical properties while obeying the symmetries of classical dynamics, such as LSC-IVR, Centroid Molecular Dynamics (CMD) and Ring Polymer Molecular Dynamics (RPMD) naturally provide short-time approximations to the multi-time symmetrized Kubo transformed correlation function. Here, we show how the third-order response can be formulated in terms of equilibrium symmetrized Kubo transformed correlation functions. We demonstrate the utility and accuracy of our approach by showing how it can be used to obtain the third-order response of a series of model systems using both classical dynamics and RPMD. In particular, we show that this approach captures features such as anharmonically induced vertical splittings and peak shifts while providing a physically transparent framework for understanding multidimensional spectroscopies.
\end{abstract}

\maketitle

\section{Introduction}

Linear spectroscopies, ranging from electronic, terahertz, Raman, SFG, and infra-red are the workhorse methods used to interrogate time and energy scales of chemical systems. However, in disordered condensed phase systems the presence of many overlapping features makes decoding the information present to obtain the individual processes and states present, the timescales of their interconversion, and the molecular motions they arise from extremely challenging. Nonlinear spectroscopies provide much richer information,\cite{Hamm2009,Jansen2019,Biswas2022} with the 2D counterparts of the aforementioned spectroscopies (2D-IR,\cite{Hamm1998,Zanni2001} 2D-Raman-Thz,\cite{Savolainen2013} 2D-Raman,\cite{Cho1999} 2D-SFG\cite{Xiong2011}) giving access to extra time and frequency dimensions that allow for easier identification of the states present and the rates of their interconversion. However, while these methods yield information on the timescales present in condensed phase chemical systems linking these to the microscopic structural changes that give rise to them often poses a significant challenge. One of the most powerful ways to achieve this link between dynamics and structure for linear spectrocopies in the condensed phase has been through the use of theory and simulation where a plethora of classical and semi-classical techniques have been introduced.\cite{Berne1967,Mukamel1985,Ahlborn2000,Lawrence2002,McRobbie2009,Liu2009,Witt2009,Liu2016,Willat2018,Trenins2019,Benson2021} However, the accurate and efficient simulation of nonlinear spectroscopies still presents significant challenges. In particular, 2D spectroscopies (2D-IR, 2D-ES, and 2D-SFG) can be simulated and understood in the impulsive limit via the third-order response function,\cite{Mukamel_Book,Cho_Book}
\begin{multline}
    R^{(3)}(t_1,t_2,t_3) = \\ \left(\frac{-i}{\hbar}\right)^3 Tr\{ \hat{D}(t_3)[\hat{C}(t_2),[\hat{B}(t_1),[\hat{A}(0),\rho_{\textrm{eq}}]]] \},
    \label{eq:R3_int}
\end{multline}
which describes how a system with an initial distribution $\rho_{\textrm{eq}}$ evolves in time following the effect of successive interactions with the experimentally applied fields (at times $0$, $t_1$, $t_2$) and the probe (at $t_{3}$) the nature of which determine the operators $\hat{A}$, $\hat{B}$, $\hat{C}$, and $\hat{D}$. These operators are determined by the particular spectroscopy of interest, for example if all operators are taken to be electronically off-resonant dipole operators then this measurement would correspond to 2D-IR. Given the insights that have been obtained into linear spectroscopies from molecular simulation and the recent advent of efficient {\it ab initio} molecular dynamics schemes\cite{Lee2007,Ceriotti2012,Marsalek2016,Marsalek2017,Sharma2021} and machine learning representations of {\it ab initio} potential energy surfaces\cite{Behler2007,Grisafi2018,Zuo2020,Chen2020,Chen2021} having an efficient and tractable route to compute the third-order response from molecular dynamics simulations would further our ability to understand and design 2D spectroscopic experiments in the condensed phase. However, unlike the single time correlation functions needed to capture linear spectroscopies if one were to expand the nested commutators to obtain a difference of correlation functions in Eq.~(\ref{eq:R3_int}) this would yield a zero third order response since one would obtain a difference of correlations functions that are all identical in the classical limit.

The lack of a straightforward approach to obtain the third-order response has spawned a number of methods to obtain it that are compatible with atomistic simulations. For cases in which only information on a particular mode is of interest (e.g. an OH or CO stretch coordinate) the use of frequency maps and cumulant approaches have proven useful\cite{}. However, when the response function is required across the entire frequency range one must resort to other methods. Many of these methods arise from the Poisson bracket method\cite{Mukamel1996,Kryvohuz2005} where the commutators in the response function are replaced with Poisson brackets, which arise from truncating the Wigner representation of the commutator keeping only the leading order term, and then replacing quantum observables with classical ones. Doing this gives a non-zero classical limit for the third-order response but requires the calculation of the stability matrix, a measure of the sensitivity of trajectories to their initial conditions, for the entire system. The elements of the stability matrix are known to grow rapidly with time making simulations increasingly difficult to converge at longer times and as the system becomes larger. 

To make the Poisson bracket based approaches more feasible finite field methods\cite{Jansen2000,Jansen2003,Yagasaki2008} have been introduced as an alternative way to compute the response. These methods apply explicit perturbations to the system to generate nonequilibrium trajectories that replace the chaotic measure with sudden kicks in the momenta. However, introducing nonequilibrium trajectories creates a parametric dependence of the applied perturbation which needs to be carefully chosen adding a further challenge to converging the results. While more recent work has shown that hybrid approaches\cite{Hasegawa2006,Yagasaki2009,Begui2022} combining nonequilibrium kicks with equilibrium trajectories can give improvements in simulation efficiency these methods remain computationally costly limiting their ability to treat large condensed phase systems especially those that involve long correlation times.

An alternative approach to obtaining the third-order response is to approximately cast it in terms of a single equilibrium multi-time correlation function multiplied by a correction factor. In this approach one first expands the commutators in Eq.~(\ref{eq:R3_int}) to yield a series of multi-time correlation functions and then obtains the correction factor by deriving connections between them in the frequency domain. The correction factor thus attempts to encode the interference between the multi-time correlation functions that appear in the nonlinear response that ultimately gives rise to the observed signal. Previous work has provided a route to achieve this for the second and third-order response functions based on the standard quantum multi-time correlation functions allowing simple classical limits to be defined.\cite{DeVane2003,DeVane2004,DeVane2005} However, the standard quantum multi-time correlation function is generally complex valued and doesn't allow for arbitrary permutations of operators. These properties of the standard correlation function, are highly undesirable as they are not satisfied by classical mechanics or semi-classical methods combining quantum statistical properties with classical-like evolution such as LSC-IVR,\cite{Wang1998} Centroid Molecular Dynamics\cite{Cao1994-2,Cao1994-4,Jang1999} (CMD) and Ring Polymer Molecular Dynamics\cite{Craig2004,Habershon2013} (RPMD), which are some of the most practical methods currently available to treat nuclear quantum effects in condensed phase systems.

For single-time correlation functions the issues associated with the classical and semi-classical approximation of the standard correlation function has been long appreciated and thus instead of approximating the standard correlation function one first recasts the property of interest in terms of the Kubo transformed correlation function\cite{Kubo1957,Zwanzig_1965} before approximating the Kubo correlation function classically or semi-classically. The symmetrized Kubo transformed correlation function\cite{Reichman2000,Jung2019} (SKTCF) extends this concept to the multi-time case since it is a quantum mechanical correlation function that possess the symmetries of a classical correlation function i.e. it is real valued, invariant to permutations of the operators and is time reversible. This makes it the natural correlation function to approximate using classical and semiclassical methods. Recent work has derived a correction factor that connects the second-order response to the two-time SKTCF \cite{Jung2018}. Since this approach only requires calculating an equilibrium two-time correlation function, which can be easily done using standard molecular dynamics software, this approach can be applied to study the second-order response in large condensed phase systems and has recently been combined with RPMD to elucidate the role of nuclear quantum effects (NQEs) in the two-dimensional Raman spectrum of liquid Neon.\cite{Tong2020} However, formulating the third-order response in terms of SKTCFs has remained a challenge.

Here we show how the third-order response can be obtained by utilizing a further generalization of the SKTCF. This allows us to express the third-order response function in the frequency domain in terms of a correction factor that can be applied to convert the equilibrium multi-time correlation functions generated by one's method of choice to the third-order response. We find that when our correction factor is combined with classical dynamics, LSC-IVR, CMD or RPMD it correctly recovers the vanishing of the third-order response in the harmonic limit for linear operators. Finally, we demonstrate the ability of our approach to capture the features present in the exact third-order response of a mildly anharmonic potential and a model of the O-H stretch of water when used with RPMD or classical dynamics.

\section{Theory}

\subsection{Symmetrized Kubo transformed correlation functions}

To begin we first introduce the framework of the SKTCF that will motivate the later approximations to the third-order response function. For transparency we consider a 1D system (the generalization to higher dimensions is straightforward) with the Hamiltonian 
\begin{equation}
H = \frac{p^2}{2m} + V(q),
\end{equation}
where $p$ and $q$ are the momentum and position. The SKTCF is\cite{Jung2019}
\begin{multline}
\label{eq:sktcf}
\langle A_0;A_1;\cdots;A_{n-1};A_n\rangle^{\textrm{sym}} = \frac{1}{\beta^nZ}\int^{\beta}_0 d\lambda_0 \int^{\beta}_0 d\lambda_1 \cdots \int^{\beta}_0 d\lambda_{n-1}  \\ \times \langle \overrightarrow{T}_{\beta} \hat{A}_0(-i\hbar\lambda_0)\hat{A}_1(-i\hbar\lambda_1+t_1) \cdots \hat{A}_{n-1}(-i\hbar\lambda_{n-1}+t_{n-1})\hat{A}_n(t_n) \rangle,
\end{multline}
where
\begin{equation}
\langle \cdots \rangle = \frac{1}{Z}Tr\left[e^{-\beta \hat{H}} \cdots \right],
\label{eq:std_corr}
\end{equation}
$Z=Tr[e^{-\beta \hat{H}}]$ is the partition function and $\hat{O}(\alpha) = e^{i\hat{H}\alpha/\hbar}\hat{O}e^{-i\hat{H}\alpha/\hbar}$ denotes the Heisenberg time evolved operators, which in this case are generally a mixture of real and imaginary time evolution. The imaginary time ordering operator $\overrightarrow{T}_{\beta}$ when applied to the imaginary time evolved operators following it gives the sum of all unique permutations of these operators and is as defined in Ref. \citenum{Jung2019}. The SKTCF possesses the properties of invariance to permutations of the operators, time reversal symmetry, and is always real valued. These are properties classical multi-time correlation functions posses as well. Appendix \ref{sec:HarmonicLimit} gives further evidence of the correspondence between the SKTCF and classical multi-time correlation functions by showing that they yield the same result for a harmonic potentials with linear operators. The standard Kubo transformed correlation,\cite{Kubo1957} which forms the basis for calculating linear spectroscopies,
\begin{equation}
\langle A;B \rangle = \frac{1}{\beta Z}\int^{\beta}_0 d\lambda \: Tr\left[ e^{-(\beta-\lambda)\hat{H}}\hat{A}(0) e^{-\lambda\hat{H}}\hat{B}(t) \right]
\end{equation}
can be viewed as a special case of Eq.~(\ref{eq:sktcf}) when $\hat{A}_{i}=\hat{1}$ (the identity operator) for $i>1$. Previous work has shown how the symmetrized double Kubo transform can be used to approximate the second-order response function\cite{Jung2018,Tong2020}
\begin{align}
&\langle A;B;C \rangle^{\textrm{sym}}=\frac{1}{\beta^2Z}\int^{\beta}_0 d\lambda \int^{\lambda}_0 d\mu \: Tr\Big[ e^{-(\beta-\lambda)\hat{H}} \nonumber  \\  &\times \left(\hat{A}(0) e^{-(\lambda-\mu)\hat{H}} \nonumber \hat{B}(t_1) + \hat{B}(t_1)e^{-(\lambda-\mu)\hat{H}}\hat{A}(0) \right)  \\  &\times e^{-\mu\hat{H}}\hat{C}(t_2) \Big].
\end{align}
Eq.~(\ref{eq:sktcf}) also reduces to this result when $\hat{A}_{i}=\hat{1}$ for $i>2$. The correlation function that will be of interest for us to derive an approximation to the third-order response function will be the three-time SKTCF
\begin{multline}
\langle A;B;C;D \rangle^{\textrm{sym}} = \langle A;B;C;D \rangle + \langle A;C;B;D \rangle + \langle C;A;B;D \rangle  \\ + \langle B;A;C;D \rangle + \langle B;C;A;D \rangle + \langle C;B;A;D \rangle  \\
 \equiv  K^{\textrm{sym}}(t_1,t_2,t_3),
\label{eq:triple_sym_Kubo}
\end{multline}
where we introduce the shorthand notation for a general triple Kubo transformed multi-time correlation function,
\begin{multline}
\langle A;B;C;D \rangle = \frac{1}{\beta^3Z}\int^{\beta}_0 d\lambda \int^{\lambda}_0 d\mu \int^{\mu}_0 d\nu \: Tr\Big[ e^{-(\beta-\lambda)\hat{H}} \\ \times \hat{A}(0) e^{-(\lambda-\mu)\hat{H}}\hat{B}(t_1)e^{-(\mu-\nu)\hat{H}}\hat{C}(t_2)e^{-\nu\hat{H}}\hat{D}(t_3) \Big],
\end{multline}
the ``;'' is used to denote the imaginary time averaging of the operator it follows where it is implied that the averaging is done in the order $\nu$, $\mu$, and then $\lambda$.

\subsection{Third-order response function}

The third-order response function is defined in Eq.~(\ref{eq:R3_int}). Here we define the initial distribution to be the equilibrium density matrix $\rho_{\textrm{eq}} = \frac{e^{-\beta \hat{H}}}{Z}$. Expanding the commutators allows one to see that the third-order response can be written in terms of standard correlation functions ($S_{i}$) as
\begin{multline}
R^{(3)}(t_1,t_2,t_3) = -
\frac{2}{\hbar^3} \textrm{Im}\Big[  S_{DCBA}(t_1,t_2,t_3) - \\ S_{CDBA}(t_1,t_2,t_3) + S_{ACDB}(t_1,t_2,t_3) - S_{ADCB}(t_1,t_2,t_3)  \Big],
\label{eq:R3_std}
\end{multline}
where
\begin{eqnarray}
\label{eq:C1}
S_{DCBA}(t_1,t_2,t_3) &=&  \langle D(t_3)C(t_2)B(t_1)A(0) \rangle, \\
S_{CDBA}(t_1,t_2,t_3) &=&  \langle C(t_2)D(t_3)B(t_1)A(0) \rangle, \\
S_{ACDB}(t_1,t_2,t_3) &=&  \langle A(0)C(t_2)D(t_3)B(t_1) \rangle, \\ 
S_{ADCB}(t_1,t_2,t_3) &=&  \langle A(0)D(t_3)C(t_2)B(t_1) \rangle,
\label{eq:C4}
\end{eqnarray}
``$\textrm{Im}$" denotes the imaginary part and the subscripts on the correlation functions denotes the ordering of the operators. The time variables $t_1,t_2,t_3$ are always associated with their respective operators $B,C,D$ (i.e. $B(t_1)$ etc.).  Obtaining a classical limit of Eq.~(\ref{eq:R3_std}) is difficult for two reasons. First, since in the classical limit all operators commute the correlation functions in Eqs.~(\ref{eq:C1})-(\ref{eq:C4}) are all the same and the third order response (Eq.~(\ref{eq:R3_std})) vanishes. Second, the standard multi-time quantum correlation functions (Eqs.~(\ref{eq:C1})-(\ref{eq:C4})) don't possess the symmetries of classical correlation functions i.e. they are complex valued and do not allow arbitrary permutation of the operators. This motivates rewriting Eq.~(\ref{eq:R3_std}) in terms of SKTCFs. In the following section we show that by doing this in the time domain one can recover the Poisson bracket method. We then show that if one works in the frequency domain it is possible to derive a generalized detailed balance correction factor that can be used to convert SKTCFs to the third-order response function.

\subsubsection{Time domain approach to expressing the third-order response in terms of SKTCFs}\label{sec:time-dom}

Working in the time domain one can express Eq.~(\ref{eq:R3_int}) in terms of SKTCFs by making repeated use of the Kubo identity\cite{Nitzan_Book}
\begin{equation}
\left 
[\hat{A},e^{-\beta\hat{H}}\right] = -i\hbar \int^{\beta}_0 d\lambda \: e^{-(\beta-\lambda)\hat{H}}\dot{\hat{A}}e^{-\lambda\hat{H}},
\label{eq:Kubo_Ident}
\end{equation}
which yields
\begin{multline}
R^{(3)}(t_{1},t_{2},t_{3}) = -\beta^3\langle \dot{A};\dot{B};\dot{C};D \rangle^{\textrm{sym}} - \frac{\beta^2}{\hbar}\Big[ 
\langle \dot{A};i[C,\dot{B}];D\rangle^{\textrm{sym}} \\ + \langle \dot{B};i[C,\dot{A}];D\rangle^{\textrm{sym}} + \langle \dot{C};i[B,\dot{A}];D\rangle^{\textrm{sym}} \Big]  \\  + \frac{\beta}{\hbar^2}\langle [C,[B,\dot{A}]];D \rangle.
\label{eq:R3_Kubo}
\end{multline}
One can take the classical limit of Eq.~(\ref{eq:R3_Kubo}) by using the classical version of the operators and replacing the commutators with Poisson brackets, $[X,Y] \to i\hbar\{X,Y\}_{\textrm{PB}}$ upon performing the integrals over $\nu$, $\mu$, and $\lambda$ the Poisson bracket approach is recovered\cite{Mukamel1996,Kryvohuz2005}
\begin{multline}
R^{(3)}(t_{1},t_{2},t_{3}) =-\beta^3\langle \dot{A}\dot{B}\dot{C}D \rangle +\beta^2 \Big[\langle \{C,\dot{A}\dot{B}\}_{\textrm{PB}}D \rangle  \\ + \langle \dot{C}\{B,\dot{A}\}_{\textrm{PB}}D \rangle \Big]  - \beta \langle \{C,\{B,\dot{A}\}_{\textrm{PB}}\}_{\textrm{PB}}D \rangle.
\end{multline}
This approach has previously been introduced and used to study water clusters and intramolecular vibrations in water.\cite{Jeon2010,Jeon2014} However, as discussed in the introduction, this method involves measures of chaos making it challenging to apply. Working with Eq.~(\ref{eq:R3_Kubo}) directly is also difficult since it requires the ability to evaluate time correlation functions of commutators. In order to remove the dependence on the commutators in the following section we move to the frequency domain where these can be avoided.

\subsubsection{Frequency domain approach to expressing the third-order response in terms of a single SKTCF} \label{sec:freq-dom}

To avoid the commutators that arise in the time domain approach to formulating the third-order response in terms of SKTCFs we can move to the frequency domain. As shown in Appendix \ref{sec:AppA} the third-order response in Eq.~(\ref{eq:R3_std}) can be transformed to the frequency domain to give
\begin{multline}
\tilde{R}^{(3)}(\omega_1,\omega_2,\omega_3) = \frac{i}{\hbar^3}\tanh(\beta\hbar\bar{\omega}/2)\Big[(\tilde{S}_{DCBA} +\tilde{S}_{ADCB})  \\ - (\tilde{S}_{CDBA} +\tilde{S}_{ACDB}) + (\tilde{S}^*_{DCBA} +\tilde{S}^*_{ADCB})  \\ - (\tilde{S}^*_{CDBA} +\tilde{S}^*_{ACDB}) \Big],
\end{multline}
where the tilde denotes a three dimensional temporal Fourier transform of a generic function, i.e. $W(t_1,t_2,t_3)$
\begin{eqnarray}
\tilde{W}(\omega_1,\omega_2,\omega_3) &=& \int^{\infty}_{-\infty} dt_1 \int^{\infty}_{-\infty} dt_2 \int^{\infty}_{-\infty} dt_3 e^{-i(\omega_1t_1 + \omega_2t_2 + \omega_3t_3)} \nonumber \\ && \times W(t_1,t_2,t_3).
\end{eqnarray}
through the further use of frequency domain relationships (see Appendix \ref{sec:AppA}) we can exactly remove the dependence on the $\tilde{S}_{ACDB}(\omega_{1},\omega_{2},\omega_{3})$ and $\tilde{S}_{ADCB}(\omega_{1},\omega_{2},\omega_{3})$ correlation functions to yield
\begin{eqnarray}
\tilde{R}^{(3)}(\omega_{1},\omega_{2},\omega_{3}) &=& \frac{i}{\hbar^3}\Big[(e^{\beta\hbar\bar{\omega}}-1)(\tilde{S}_{DCBA} - \tilde{S}_{CDBA}) - \nonumber \\ && (e^{-\beta\hbar\bar{\omega}}-1)(\tilde{S}^*_{DCBA} - \tilde{S}^*_{CDBA}) \Big],
\label{eq:R3_freqq}
\end{eqnarray}
where $\bar{\omega}=\omega_1+\omega_2+\omega_3$. Unlike the expressions previously derived for the second-order response function\cite{Jung2018} where the response only depended on a single two-time correlation function the expression for the third-order response involves a difference of two correlation functions necessitating further approximation to obtain an expression that is tractable to be combined with classical-like methods. 

Up to this point all the manipulations we have performed are exact. However, since classical correlation functions are agnostic to the ordering of operators we want to obtain an expression that expresses the third-order response in terms of a single correlation function rather than a difference of two correlation functions to avoid it vanishing. To achieve this one can show that in the harmonic limit ($V(q)=\frac{m\Omega^2}{2}q^2$) with linear operators $\tilde{S}_{DCBA} = g(\omega_1,\omega_2,\omega_3) \tilde{S}_{CDBA}$\cite{DeVane2004} where
\begin{equation}
g(\omega_1,\omega_2,\omega_3) = \frac{1+e^{-\beta\hbar(\omega_1+\omega_3)/2}}{1+e^{-\beta\hbar(\omega_1+\omega_2)/2}}.
\label{eq:AtoB}
\end{equation}
Note that in what follows we use lowercase letters to denote frequency dependent factors while uppercase letters are used for time correlation functions. Using the relationship in Eq.~(\ref{eq:AtoB}) to relate $\tilde{S}_{DCBA}$ and $\tilde{S}_{CDBA}$ gives
\begin{multline}
\tilde{R}^{(3)}(\omega_{1},\omega_{2},\omega_{3}) = \frac{i}{\hbar^3}\Big[(e^{\beta\hbar\bar{\omega}}-1)(g(\omega_{1},\omega_{2},\omega_{3})-1)\tilde{S}  \\  - (e^{-\beta\hbar\bar{\omega}}-1)(g(-\omega_{1},-\omega_{2},-\omega_{3})-1)\tilde{S}^* \Big],
\label{eq:R3_firstapprox}
\end{multline}
where we have defined $\tilde{S}(\omega_{1},\omega_{2},\omega_{3})=\tilde{S}_{CDBA}(\omega_{1},\omega_{2},\omega_{3})$ since it is now the only correlation function involved in obtaining the third-order response. 

We can then derive (see Appendix \ref{sec:AppB}) and use the exact relation between the three-time Kubo transformed correlation function and the standard correlation function in frequency space
\begin{equation}
\tilde{S}(\omega_1,\omega_2,\omega_3) = \frac{\tilde{K}(\omega_1,\omega_2,\omega_3)}{f(\omega_1,\omega_2,\omega_3)},
\label{eq:StoK}
\end{equation}
and
\begin{equation}
\tilde{S}^*(\omega_1,\omega_2,\omega_3) = \frac{\tilde{K}^*(\omega_1,\omega_2,\omega_3)}{f(-\omega_1,-\omega_2,-\omega_3)},
\label{eq:StoK_neg}
\end{equation}
to obtain an expression for the third order response in terms of Kubo transformed correlation functions
\begin{multline}
\tilde{R}^{(3)}(\omega_1,\omega_2,\omega_3) = i\Big[a(\omega_1,\omega_2,\omega_3)\tilde{K}(\omega_1,\omega_2,\omega_3)  \\ - a(-\omega_1,-\omega_2,-\omega_3)\tilde{K}^*(\omega_1,\omega_2,\omega_3) \Big],
\label{eq:R3_babyKubo}
\end{multline}
where $K = \langle B;A;C;D \rangle$ is the three-time Kubo transformed correlation function and $\tilde{K}$ is its three dimensional temporal Fourier transform with
\begin{equation}
a(\omega_1,\omega_2,\omega_3) = \frac{[e^{\beta\hbar\bar{\omega}}-1][g(\omega_1,\omega_2,\omega_3)-1]}{\hbar^3f(\omega_1,\omega_2,\omega_3)},
\end{equation}
and
\begin{eqnarray}
f(\omega_1,\omega_2,\omega_3) &=& \frac{1}{\beta^3\hbar^3}\Bigg[ \frac{e^{\beta\hbar\bar{\omega}}-1}{\omega_1\bar{\omega}(\omega_1+\omega_3)} 
+ \frac{e^{\beta\hbar\omega_2}-1}{\omega_2\omega_3(\omega_1+\omega_3)} \nonumber \\ &&
- \frac{e^{\beta\hbar(\omega_2+\omega_3)}-1}{\omega_1\omega_3(\omega_2+\omega_3)}
\Bigg].
\label{eq:f_factor_intform}
\end{eqnarray}
In the SM Sec.~I we demonstrate the analytic properties of Eq.~(\ref{eq:f_factor_intform}).

\begin{figure}[h]
    \begin{center}
        \includegraphics[width=0.5\textwidth]{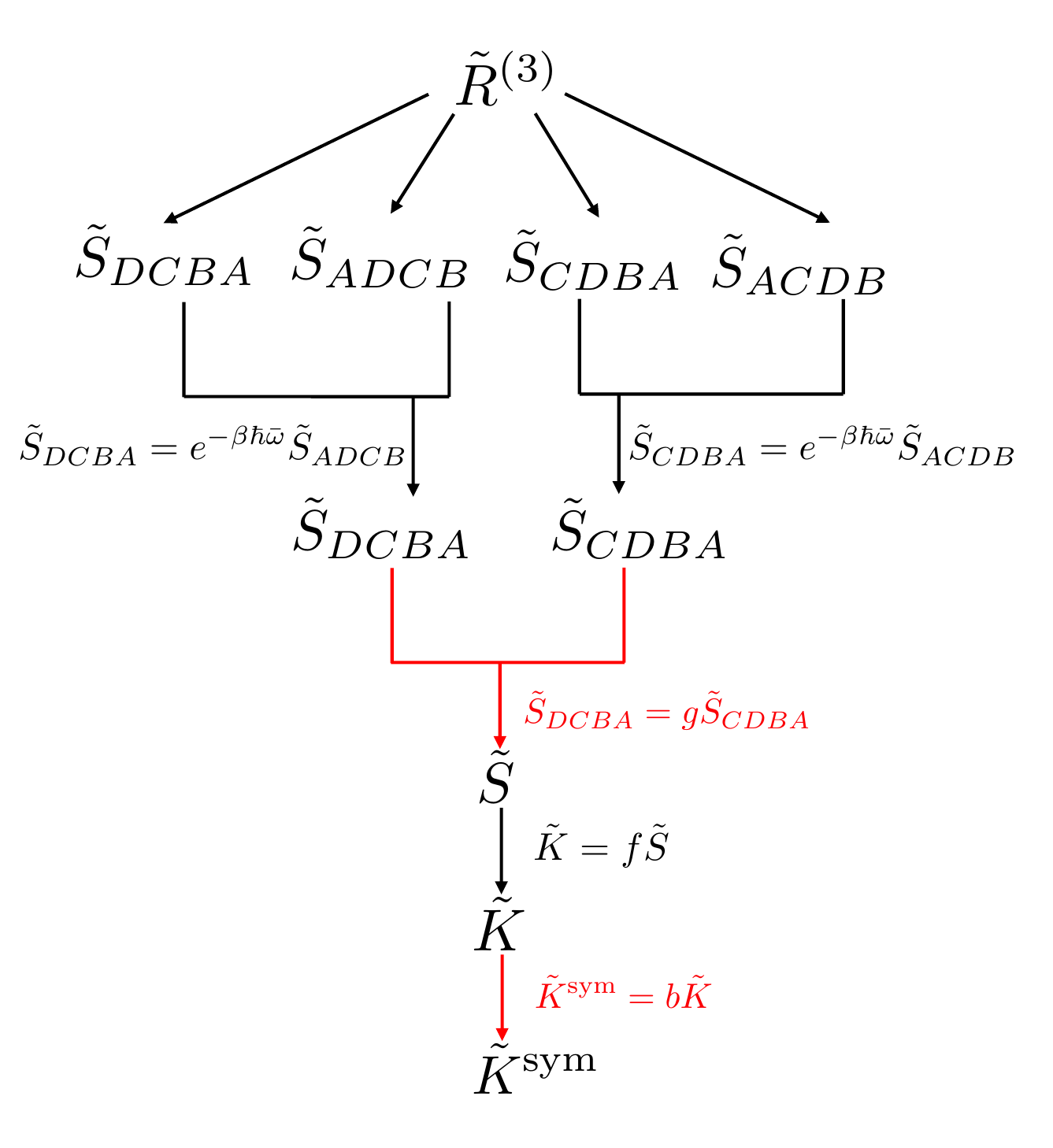}
    \end{center}
    \vspace{-8mm}
    \caption{Diagram depicting the mathematical relationships used in Sec.~\ref{sec:freq-dom} to connect the correlation functions required to compute the third-order response function ($\tilde{R}^{(3)}$) to the SKTCF ($\tilde{K}^{sym}$). Black lines denote exact relationships while red lines are only formally satisfied in the harmonic limit for linear operators.}
	\label{fig:Corr_sum}
\end{figure}

At this stage it is worth noting some desirable features of Eq.~(\ref{eq:R3_babyKubo}). Whereas Eq.~(\ref{eq:R3_firstapprox}) has a prefactor of $1/\hbar^3$ by converting from the standard to the Kubo transformed correlation functions the factor of $\hbar^{3}$ that appears in the denominator of $f(\omega_1,\omega_2,\omega_3)$ cancels that in the denominator of $a(\omega_1,\omega_2,\omega_3)$ thus ensuring a well defined classical ($\hbar\to0$) limit. Additionally, even though we made one harmonic approximation to derive the connection used in Eq.~(\ref{eq:R3_firstapprox}) the overall expression we have derived for the third-order response still retains its overall odd symmetry with respect to the frequency variables, which can be seen by noting that from the general properties of Fourier transforms $\tilde{K}^*(\omega_1,\omega_2,\omega_3)=\tilde{K}(-\omega_1,-\omega_2,-\omega_3)$ and inserting this into Eq.~(\ref{eq:R3_babyKubo}). This symmetry is important to capture the physical properties of signals governed by the third-order response by ensuring the response averaged over all times is zero.

To obtain a third-order response expression that is in terms of the Fourier transformed SKTCF $\tilde{K}^{\textrm{sym}}(\omega_1,\omega_2,\omega_3)$, which is a real quantity, we now separate $\tilde{K}(\omega_1,\omega_2,\omega_3)$ and $\tilde{K}^*(\omega_1,\omega_2,\omega_3)$ into their even and odd components and exploit the fact that in the harmonic limit the odd component vanishes (see SM Sec. II for proof) to obtain the third order response purely in terms of the real part of the correlation function $\textrm{Re}[\tilde{K}]$, 
\begin{multline}
\tilde{R}^{(3)}(\omega_1,\omega_2,\omega_3) = \\ -i[a(\omega_1,\omega_2,\omega_3)-a(-\omega_1,-\omega_2,-\omega_3)]\textrm{Re}[\tilde{K}].
\end{multline}
To express the response in terms of the SKTCF we utilize one final harmonic approximation to relate $\tilde{K}^{\textrm{sym}}(\omega_1,\omega_2,\omega_3)$ and $\textrm{Re}[\tilde{K}(\omega_1,\omega_2,\omega_3)]$ (see SM Sec. II) to arrive at our main result
\begin{equation}
\label{eq:Kubo_response1}
\tilde{R}^{(3)} (\omega_1,\omega_2,\omega_3) = -ih(\omega_1,\omega_2,\omega_3) \tilde{K}^{sym}(\omega_1,\omega_2,\omega_3),
\end{equation}
where
\begin{multline}
\label{eq:QFC_Kubo}
h(\omega_1,\omega_2,\omega_3) = \\ b(\omega_1,\omega_2,\omega_3)[a(\omega_1,\omega_2,\omega_3)-a(-\omega_1,-\omega_2,-\omega_3)].
\end{multline}
and
\begin{equation}
b(\omega_1,\omega_2,\omega_3) = \frac{\sinh(\beta\hbar\bar{\omega}) - \beta\hbar\bar{\omega}}{\beta\hbar\bar{\omega}[\cosh(\beta\hbar\bar{\omega})-1]}.
\label{eq:K_to_Ksym}
\end{equation}
The correction factor we have derived in Eq.~(\ref{eq:QFC_Kubo}) guarantees the response vanishes when $\bar{\omega}=0$ which is in agreement with the exact response function. When the SKTCF is computed either classically, with LSC-IVR, CMD or RPMD in the harmonic limit with linear operators it also gives the correct limit of zero response. The function defined in Eq.~(\ref{eq:K_to_Ksym}) is non-negative and serves as a scaling factor for the intensities as it cannot change the sign or zero out any peaks in the frequency domain. It is also an even function and thus keeps the overall odd symmetry of the response. Figure~\ref{fig:Corr_sum} outlines the mathematical connections that were used in our derivation with the relationships shown as black lines being exact and those as shown as red lines only being formally satisfied in the harmonic limit for linear operators.

The limit as $\hbar\to0$, i.e. the classical limit of the correction factor is
\begin{equation}
\label{eq:QCF_CL}
h^{\textrm{cl}}(\omega_1,\omega_2,\omega_3) =\lim_{\hbar\to0} h(\omega_1,\omega_2,\omega_3) = \frac{\beta^3}{4}\bar{\omega}(\omega^2_3-\omega^2_2).
\end{equation}
One key advantage of the classical limit of the correction factor is that it can be applied in the time domain since it can be analytically inverse Fourier transformed yielding a sequence of derivatives. An alternative correction factor shown in Appendix \ref{sec:AppC} can be obtained by starting from Eq.~(\ref{eq:R3_firstapprox}) and rather than moving to the Kubo transformed correlation functions one instead performs the frequency space connections between the standard quantum correlation functions in the harmonic limit. This approach was first described in Ref.~\citenum{DeVane2004} and despite the two approaches yielding different correction factors the form of the classical limit of each factor is the same. Connecting the third-order response to the standard quantum correlation function leads to a different scaling in $\hbar$ for the real and imaginary components of the standard correlation function which necessitates the use of an approximate relationship between the real and imaginary components of the correlation function so as to give a consistent scaling of $\hbar$ to the entire response function\cite{DeVane2004}. The use of this approximate connection is avoided when connecting the third-order response to the SKTCF, since as shown in Eq.~(\ref{eq:R3_babyKubo}) the $1/\hbar^3$ prefactor inherent to the response function naturally cancels due to the thermal factor introduced by the multi-time Kubo transform (see Eq.~(\ref{eq:f_factor_intform})).

Finally, we note that in the derivations above, for mathematical convenience, we use the absolute time convention ($t_{1},t_{2},t_{3}$). 2D experiments are typically often analyzed in the relative time convention ($t_{1},t_{1}+t_{2},t_{1}+t_{2}+t_{3}$). To convert between the two one makes a simple change of variables of $t_1 \to t'_1$, $t_2\to t'_1+t'_2$, and $t_3\to t'_1+t'_2+t'_3$ where the primed variables are those given in the relative convention. This change of time variables changes the frequency variables to $\bar{\omega}\to \omega'_1$, $\omega_2 + \omega_3 \to \omega'_2$, and $\omega_3 \to \omega'_3$.

\begin{figure*}[ht]
    \begin{center}
        \includegraphics[width=0.8\textwidth]{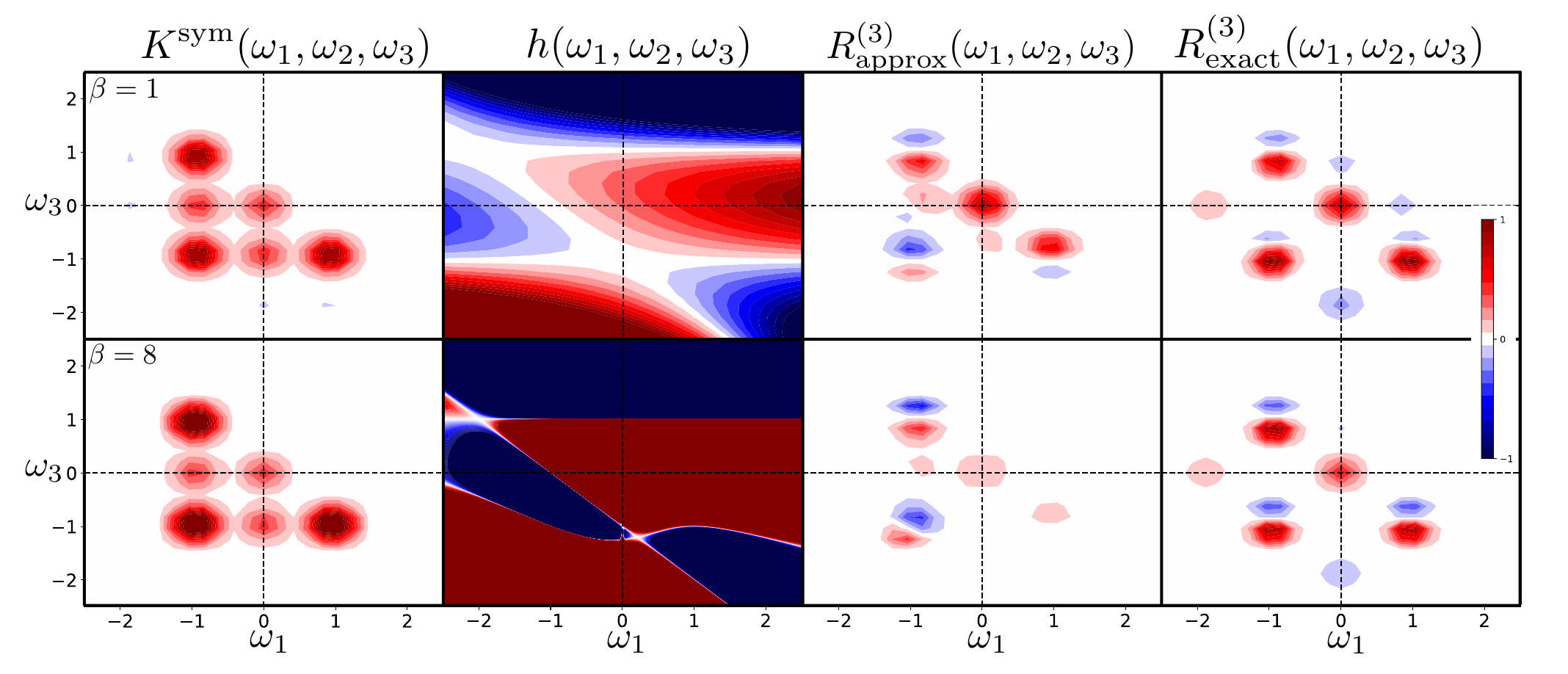}
    \end{center}
    \vspace{-5mm}
    \caption{Construction of the third-order response for the MAP model from the SKTCF according to Eq.~(\ref{eq:Kubo_response1}) at inverse temperatures $\beta=1$ (top row) and $\beta=8$ (bottom row) for $\omega_2=1$. The first column shows the 3D Fourier transform of the SKTCF calculated using exact quantum dynamics. The second column shows our correction factor at that temperature. The third column shows approximate third-order response obtained from multiplying the exact SKTCF (first column) by our correction factor (second column). This approximation to the third-order response can be compared to that computed exactly in the fourth column obtained from Eq.~(\ref{eq:R3_std}). Since the SKTCF is computed exactly all the error in comparing the third-order response in columns three and four arises from inaccuracies in our correction factor for the MAP model.}
	\label{fig:Res_demo}
\end{figure*}

\section{Computational Details}
\label{sec:CompDets}

To test the accuracy of our correction factor and its classical limit when combined with the exact and approximate SKCTF we performed simulations of two systems of differing levels of anharmonicity. The first is a mildly anharmonic potential (MAP) that has been used in a number of previously studies to benchmark the performance of dynamics methods for single-time correlation functions\cite{Jang1999,Craig2004,Perez2009}. To provide a more challenging test we also consider a O-H stretch potential (OHP) that is more strongly anharmonic potential where the parameters and functional form were based on those from the qTIP4P/F O-H stretch potential\cite{Habershon2009}. In atomic units the the MAP model is:
\begin{equation}
V(q)=\frac{1}{2} q^2 + \frac{1}{10} q^3 + \frac{1}{100} q^4
\label{eq:MAP}
\end{equation}
and the OHP is
\begin{equation}
V(q) = \frac{1}{2} q^2 - \frac{\alpha}{2} q^3 + \frac{7\alpha^2}{24} q^4
\label{eq:OHP}
\end{equation}
with $\alpha= 1.21$. For both of these systems we performed simulations at an inverse temperature of $\beta=8$, where only the ground state is appreciably populated, and $\beta=1$, where states up to the fifth vibrational level have significant population in both models.

For each of these models we applied our correction factor to the exact SKTCF and the TRPMD\cite{Rossi2014,Hele2016} and classical approximations to the SKTCF and compared them to the exact third-order response function. The exact calculations of the SKTCF and third-order response function were performed by first constructing the Hamiltonian in a discrete variable representation\cite{Marston1989} comprised of 1550 grid points, diagonalizing the Hamiltonian and then evaluating the trace in the energy-eigenbasis using the lowest eight eigenvalues and their corresponding eigenvectors. The TRPMD and classical approximations of the SKTCF\cite{Hele2015,Hele2016,Jung2020} were obtained using the same settings employed in Ref.~\citenum{Jung2018} with $32$ beads used in all of the TRPMD simulations.

All of the simulations of the correlation functions were performed in the time domain. A multi-dimensional Hann window was applied to these time domain functions such that they decayed to zero at $t_1,t_2,t_3=\pm15$ and then the FFT algorithm was used to perform the multi-dimensional Fourier transforms.

\begin{figure*}[]
    \begin{center}
        \includegraphics[width=0.8\textwidth]{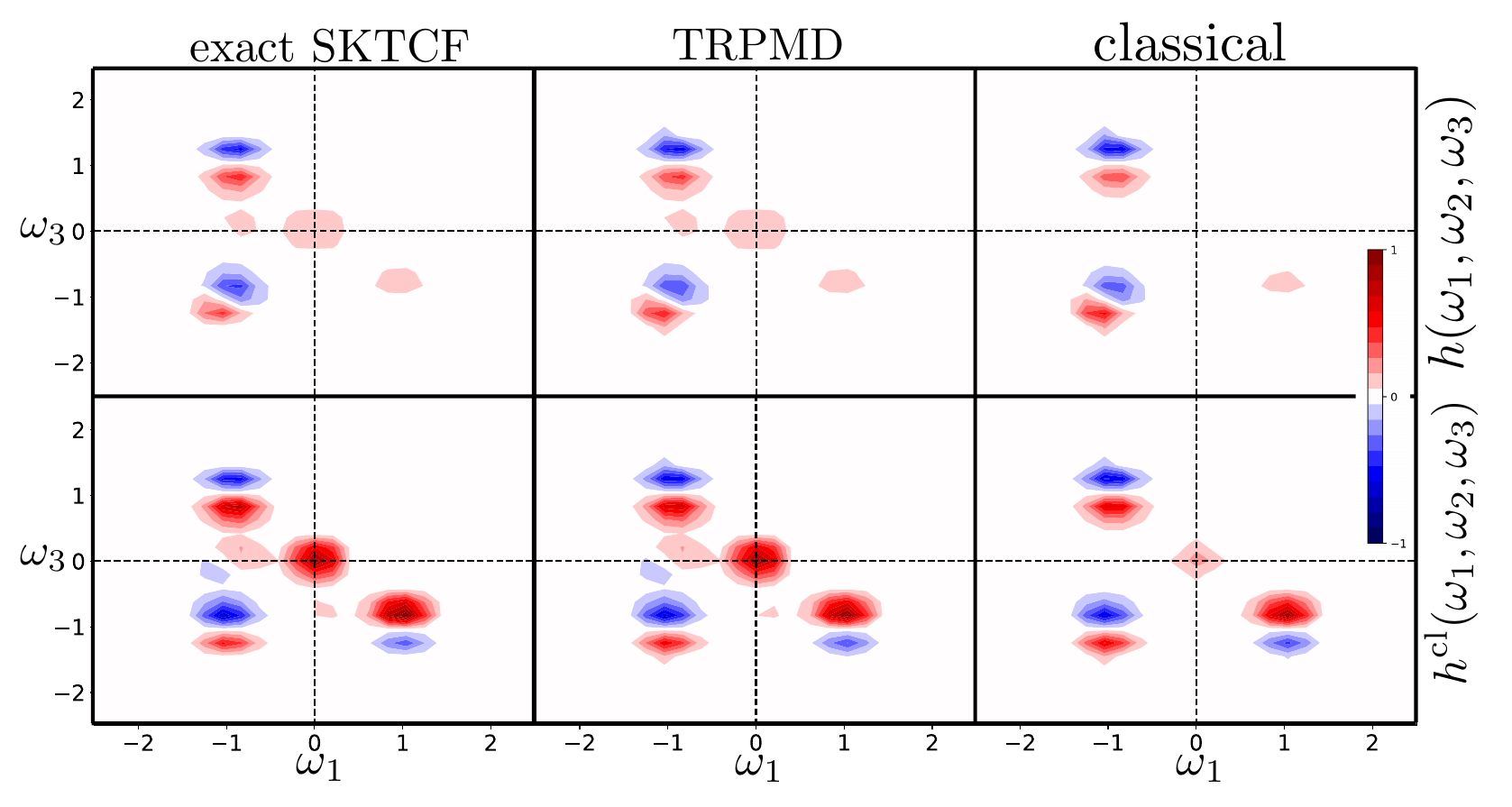}
    \end{center}
    \vspace{-5mm}
    \caption{Third-order response obtained for the MAP model at $\beta=8$ and $\omega_2=1$ using the full correction factor (top row) and its classical limit (bottom row). The columns show the effect of three levels of dynamics used to compute the SKTCF to which the correction factor is applied to yield the third-order response: exact quantum dynamics (first column), TRPMD (second column), and classical dynamics (third column).}
	\label{fig:MAP_freqcompare}
\end{figure*}

\section{Results and Discussion}

To test the approximations made in deriving our correction factor to convert the SKTCF into the third-order response given in Eq.~(\ref{eq:Kubo_response1}) we applied it to SKTCFs obtained from exact and approximate (TRPMD and classical) dynamics for the two potentials of varying anharmonicity, denoted MAP and OHP, defined in Eqs.~(\ref{eq:MAP}) and (\ref{eq:OHP}). All operators are taken to be linear position operators (i.e. $\hat{A}=\hat{B}=\hat{C}=\hat{D}=\hat{q}$). With this choice of operators a purely harmonic potential would yield a zero third-order response which would be exactly captured by our correction factor with RPMD, exact or classical dynamics. Hence the responses probed in the following section leading to peak splittings arise purely from the anharmonic nature of the potential energy.

We first present the effects of applying the correction factor in Eq.~(\ref{eq:QFC_Kubo}) to the exact three-time SKTCF in Fig.~\ref{fig:Res_demo} for both the low and high temperature regimes of the MAP.  By comparing the left hand panels of Fig.~\ref{fig:Res_demo} which shows the exact SKTCF, $K^{\textrm{sym}}(\omega_1,\omega_2,\omega_3) $, to the right hand panels which show the exact 3rd order response $R^{(3)}_{\textrm{exact}}(\omega_1,\omega_2,\omega_3)$  at $\beta=1$ and $8$ it is clear that the SKTCF itself does not resemble the response function. In particular the SKTCF only contains positive features while the third-order response contains both positive and negative ones. Applying our correction factor, $h(\omega_1,\omega_2,\omega_3)$, (middle-left panel) derived in Sec.~\ref{sec:freq-dom} to the exact SKTCF yields the approximate third-order response  $R^{(3)}_{\textrm{approx}}(\omega_1,\omega_2,\omega_3)$ shown in the middle-right panel. With the correction factor applied to the SKTCF the characteristic positive-negative splittings that arise from the anharmonicity of the potential emerge. The role that the correction factor plays in higher order spectroscopies is thus distinct from that in linear spectroscopy in that it no longer simply modulates intensities, but can additionally change the sign of peaks.

Figure \ref{fig:MAP_freqcompare} compares the effects on the third-order response of applying the full correction factor and its classical limit, $h^{cl}(\omega_1,\omega_2,\omega_3)$, given in Eq.~(\ref{eq:QCF_CL}) on differing levels of dynamical approximations to the SKTCF in the low temperature regime. In comparing the left panels with the middle panels it can be seen that using the SKTCF obtained from TRPMD rather than the exact SKTCF to compute the third-order response in this regime introduces virtually no extra errors on top of those already arising from the correction factors $h(\omega_1,\omega_2,\omega_3)$ and $h^{\textrm{cl}}(\omega_1,\omega_2,\omega_3)$. In contrast, classical dynamics fails to capture some features and underestimates others such as the peak at $(\omega_1=0,\omega_2=1,\omega_3=0)$ and the feature at $(\omega_1=-1,\omega_2=1,\omega_3=0)$. Comparing the full correction factor to its classical limit (top row vs. bottom row) the main difference observed is for the feature near $(\omega_1=1,\omega_2=1,\omega_3=-1)$ which is eliminated by the full correction factor but not by its classical limit. Comparing these results to the exact third order response for this system (Fig.~\ref{fig:Res_demo} bottom-right panel) one observes that this feature is present in the exact result but inverted in sign compared to what is predicted by the classical limit of the correction factor and hence both the full and classical correction factors are incorrectly capturing this feature. The results of applying the standard quantum correlation function correction factor introduced in Ref.~\citenum{DeVane2004} and derived in Appendix~\ref{sec:AppC} to the exact and approximate SKTCF correlations functions is shown in the SM and is observed to perform worse than the full correction factor introduced here or its classical limit.

\begin{figure*}[]
    \begin{center}
        \includegraphics[width=0.8\textwidth]{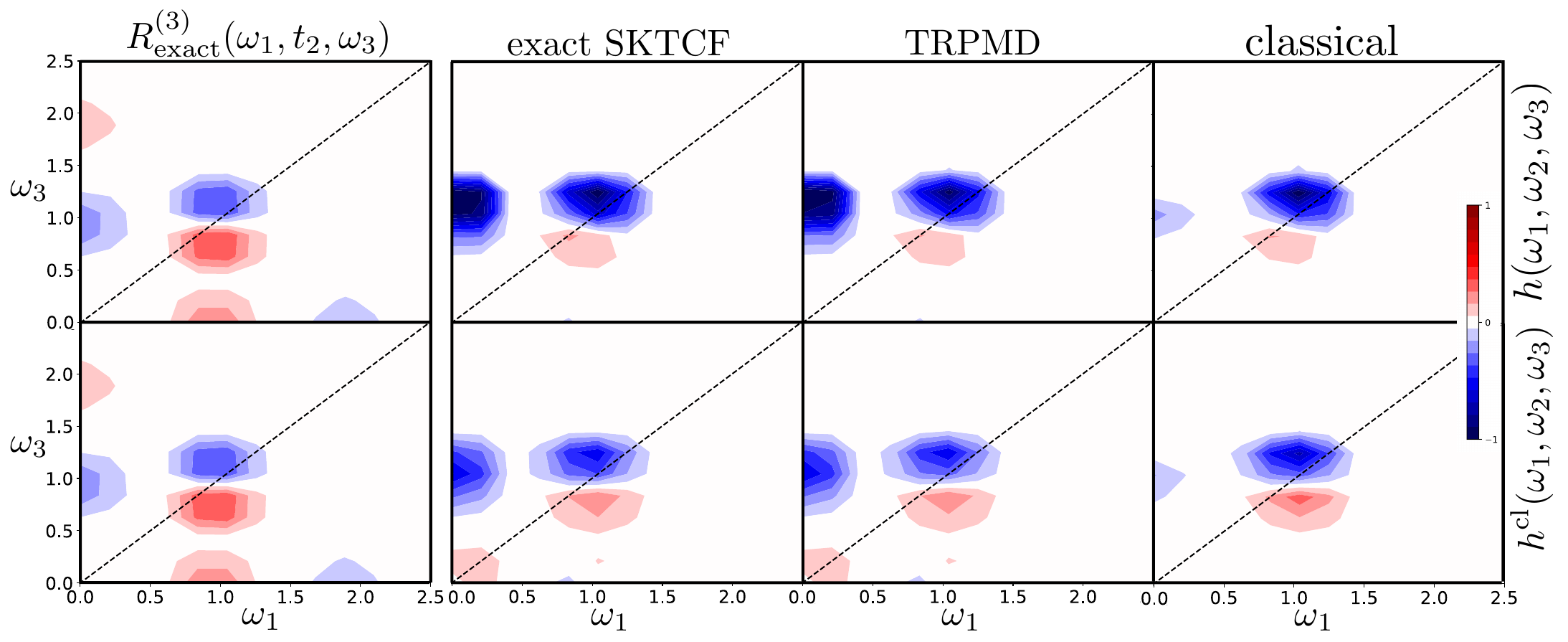}
    \end{center}
    \vspace{-5mm}
    \caption{Third-order order response for the MAP model with the $\omega_2$ axis transformed to the time domain, $R^{(3)}(\omega_1,t_2,\omega_3)$ at $t_2=0$ and $\beta=8$. The top row uses the full correction factor while the bottom row uses its classical limit. The columns correspond to the levels of dynamics used to which the correction factor is applied to yield the third-order response: exact quantum dynamics (first column), TRPMD (second column), and classical dynamics (third column).}
	\label{fig:MAP_timecompare}
\end{figure*}

So far we have demonstrated the performance of our formalism in capturing the response function entirely in the frequency domain, $R^{(3)}(\omega_1,\omega_2,\omega_3)$. However, spectroscopic measurements are frequently reported as a function of the population time delay, $t_2$, i.e as $R^{(3)}(\omega_1,t_2,\omega_3)$. To assess how the errors introduced by the correction factors manifest in this form of the third-order response Fig.~\ref{fig:MAP_timecompare} compares $R^{(3)}(\omega_1,t_2,\omega_3)$ computed exactly to the results obtained using the exact SKTCF and applying our correction factor as well as using TRPMD and classical approximations to the SKTCF. In all cases the main anharmonic splitting feature centered around ($\omega_1=1,t_2=0,\omega_3=1$) is captured by all of the dynamics methods used as well as the feature at ($\omega_1=0,t_2=0,\omega_3=1$). We show the results only at $t_2=0$ since the model is one dimensional and hence, due to the lack of interaction with other degrees of freedom, there are no new features that appear at later time delays.

By comparing the top and bottoms panels of Fig.~\ref{fig:MAP_timecompare} with the former using the full correction factor, $h(\omega_1,\omega_2,\omega_3)$, and the latter its classical limit, $h^{\textrm{cl}}(\omega_1,\omega_2,\omega_3)$, one sees that while both capture the main anharmonic feature the full correction factor gives a better description of the vanishing signal at $(\omega_1=0,t_2=0,\omega_3=0)$. While some of the features present in exact signal (e.g. those at $(\omega_1=1.8,t_2=0,\omega_3=0)$ and $(\omega_1=0,t_2=0,\omega_3=1.8)$) are not captured by any of the dynamics approaches (including the exact SKTCF) when combined with our correction factor it is remarkable that given that the connections used to derive it invoked harmonic limits that the dominant anharmonic splitting centered around $(\omega_1=1,t_2=0,\omega_3=1)$ which consists of a negative lobe above the diagonal and a corresponding positive one below it is captured. These positive and negative regions arise from the smaller spacings of the energy levels of the anharmonic potential such that the bleach and stimulated emission peaks appear shifted and opposite in sign to the excited state absorption. The ability of even classical dynamics combined with the classical limit of our correction factor to capture these physics is thus notable. The results for the high temperature regime are shown in SM Figs.~2 and 3 and exhibit similar features. 

\begin{figure}[]
    \begin{center}
        \includegraphics[width=0.45\textwidth]{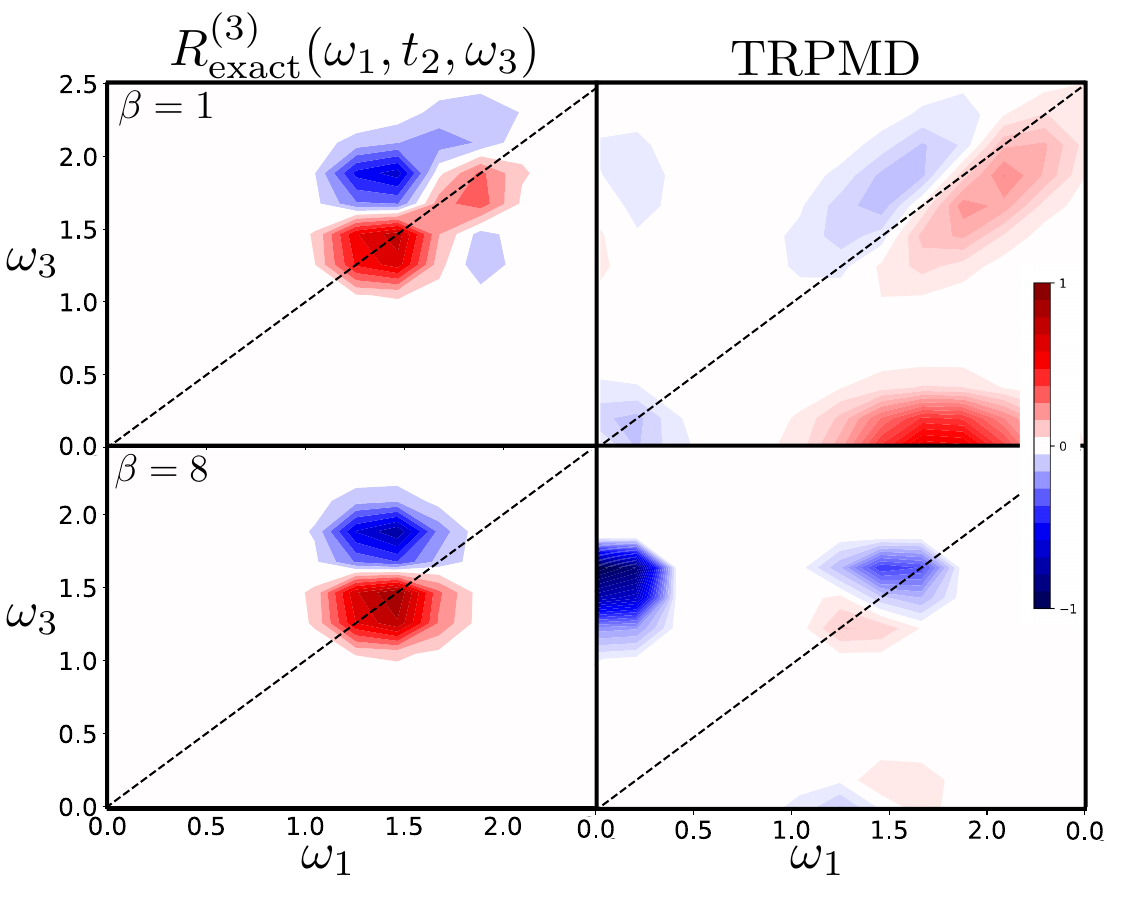}
    \end{center}
    \vspace{-5mm}
    \caption{Third-order response for the more strongly anharmonic OHP model at $\beta=1$ (top row) and $\beta=8$ (bottom row) at $t_2=0$. The left column contains the exact third-order response while the right column shows the third-order response obtained using the TRPMD approximated SKTCF in conjunction with the full correction factor.}
    \label{fig:OHP_timecompare}
\end{figure}

Finally, to test the performance of our correction factor when pushed outside the limits it was derived in we consider the third-order response ($R^{(3)}(\omega_1,t_2,\omega_3)$) in the more anharmonic OHP potential. Fig.~\ref{fig:OHP_timecompare} shows that even with this increased anharmonicity using the TRPMD SKTCF combined with our correction factor one is again able to capture the main positive and negative features split about the diagonal which for this system in the higher temperature case ($\beta=1$) gives rise to an extended oblong feature arising from overlapping transitions. However, as one would expect given the more anharmonic potential the spurious peaks at $(\omega_1=1.5,t_2=0,\omega_3=0)$ and $(\omega_1=0,t_2=0,\omega_3=1.5)$ have much higher intensity than they did in the MAP case due to the correction factors not properly dampening these peaks in spectral representation of the SKTCF, and similar failures can also be observed in the all frequency representation of the third-order response, $R^{(3)}(\omega_1,\omega_2,\omega_3)$, shown in SM Fig.~4. The comparison using the classical limit of the correction factor is shown in SM Fig.~5.

\section{Conclusion}

Using the formalism of the SKTCF we have provided a tractable approach to compute third-response functions from equilibrium trajectory based methods. The ability to do this allows for the simulation of spectroscopies that arise from third-order response, such as 2D-IR, from trajectories obtained using methods such as TRPMD, CMD, LSC-IVR or classical molecular dynamics without having to perform additional simulations with explicit field perturbations or nonequilibrium trajectories. Our correction factor is able to exactly capture the zero third-order response in the harmonic limit and despite invoking harmonic connections between correlation functions it can be combined with TRPMD or classical dynamics to reproduce the positive-negative splittings present in the third-order response of anharmonic systems which are a hallmark of higher order spectroscopies. These positive and negative splittings, which are not captured by simpler 2D correlation methods\cite{Noda2009,Morawietz2018}, arise from the action of our correction factor on the SKTCF and hence in these higher order spectroscopies the correction factor plays a more crucial role than in linear spectroscopies where it modulates intensities and does not lead to sign changes. There are limits on the success one can expect from such an approach. As demonstrated in our results pushing the current correction factor into highly anharmonic regimes leads to a notable degradation in the quality of the results obtained. The current strategy however provides a tractable approach to utilize the latest developments in efficient ab initio path integral methods\cite{Markland2018} and machine learned potentials to describe the nonlinear spectroscopy of complex systems that have recently been studied experimentally by 2D-IR ranging from protonated water\cite{Thmer2015} to ionic solutions\cite{Yuan2019}.

\section*{Acknowledgments}
This work was supported by National Science Foundation Grant No. CHE-2154291.

\section*{Supplementary Material}
The supplementary material includes details of the limiting cases of Eq.~(\ref{eq:f_factor_intform}), a derivation of Eq.~(\ref{eq:K_to_Ksym}) and additional figures demonstrating the performance of the various response function approximation schemes.

\section*{Data Availability}
The data that support the findings of this study are contained within the article and its supplementary material.

\appendix

\section{Harmonic Limit of $K^{sym}(t_1,t_2,t_3)$ for linear operators}
\label{sec:HarmonicLimit}
In the harmonic limit one can analytically evaluate the symmetrized Kubo transform. While the expressions are generally unwieldy when nonlinear operators are involved it gives a simple expression for linear position operators, namely
\begin{multline}
    K^{sym}(t_1,t_2,t_3) = \frac{1}{(\beta m \Omega^2)^2}\Big\{ \cos[\Omega(t_1-t_2-t_3)] + \\ \cos[\Omega(t_1+t_2-t_3)]  + \cos[\Omega(t_1-t_2+t_3)] 
    \Big\},
\end{multline}
which is identical to the result one would obtain for the classical correlation $\langle q(0)q(t_1)q(t_2)q(t_3) \rangle$. This is analgous to the relationship that is found in the single time case i.e., the Kubo transformed correlation function and the classical correlation function of linear position operators both give the same result of
\begin{equation}
    K(t) = \frac{1}{\beta m \Omega^2}\cos(\Omega t).
\end{equation}
This behavior is in general true for any SKTCF of the harmonic oscillator involving linear operators reinforcing the idea that the SKTCF behaves more similar to its classical counterpart than other forms of multi-time correlation functions. When nonlinear operators are involved the agreement between the SKTCF and classical ones no longer holds.
The RPMD and CMD approximations to the symmetrized Kubo transform are also exact in this limit since in the case of linear operators only the centroid of the ring polymer contributes to the correlation function yielding the classical result.

\section{Fourier relationships between standard three-time correlation functions}
\label{sec:AppA}

Each of the standard correlation functions defined in the response function can be expanded in the eigenbasis of the Hamiltonian and then subsequently Fourier transformed. The results are as follows:
\begin{eqnarray}
\tilde{S}_{DCBA}(\omega_1,\omega_2,\omega_3) &=& \frac{1}{Z}\sum_{klmn}e^{-\beta E_k} D_{kl}C_{lm}B_{mn}A_{nk} \label{eq:C1_FT}  \\ && \times\delta(\omega_1-\omega_{mn})\delta(\omega_2-\omega_{lm})\delta(\omega_3-\omega_{kl}), \nonumber \\
\tilde{S}_{CDBA}(\omega_1,\omega_2,\omega_3) &=&  \frac{1}{Z}\sum_{klmn}e^{-\beta E_k} C_{kl}D_{lm}B_{mn}A_{nk} \\ && \times\delta(\omega_1-\omega_{mn})\delta(\omega_2-\omega_{kl})\delta(\omega_3-\omega_{lm}), \nonumber \\
\tilde{S}_{ACDB}(\omega_1,\omega_2,\omega_3) &=&  \frac{1}{Z}\sum_{klmn}e^{-\beta E_k} A_{kl}C_{lm}D_{mn}B_{nk} \\ && \times\delta(\omega_1-\omega_{nk})\delta(\omega_2-\omega_{lm})\delta(\omega_3-\omega_{mn}),\nonumber \\ 
\tilde{S}_{ADCB}(\omega_1,\omega_2,\omega_3) &=&  \frac{1}{Z}\sum_{klmn}e^{-\beta E_k} A_{kl}D_{lm}C_{mn}B_{nk} \\ && \times\delta(\omega_1-\omega_{nk})\delta(\omega_2-\omega_{mn})\delta(\omega_3-\omega_{lm}),\nonumber
\label{eq:C4_FT}
\end{eqnarray}
where $\omega_{kl} = (E_k-E_l)/\hbar$ and $O_{kl} = \langle k|\hat{O} |l \rangle $. Enforcing the delta functions in Eq.~(\ref{eq:C1_FT}) we obtain
\begin{eqnarray}
\tilde{S}_{DCBA} &=&\frac{e^{-\beta\hbar\bar{\omega}}}{Z} \sum_{klmn}e^{-\beta E_n} D_{kl}C_{lm}B_{mn}A_{nk}\nonumber \\ && \times\delta(\omega_1-\omega_{mn})\delta(\omega_2-\omega_{lm})\delta(\omega_3-\omega_{kl}), \nonumber \\
&=& e^{-\beta\hbar\bar{\omega}}\tilde{S}_{ADCB}.
\end{eqnarray}
Similarly one can also show
\begin{equation}
\tilde{S}_{CDBA} = e^{-\beta\hbar\bar{\omega}}\tilde{S}_{ACDB}.
\end{equation}
Using the fact that
\begin{multline}
\tilde{W}(-\omega_1,-\omega_2,-\omega_3) = \int^{\infty}_{\infty} dt_1 \int^{\infty}_{\infty} dt_2 \int^{\infty}_{\infty} dt_3  \\ \times e^{-i(\omega_1t_1 + \omega_2t_2 + \omega_3t_3)}  W^*(t_1,t_2,t_3)
\label{eq:neg_freq}
\end{multline}
i.e. the Fourier transform of the complex conjugate gives the same frequency space function but with negative arguments we immediately get
\begin{eqnarray}
\tilde{S}^*_{DCBA} = e^{\beta\hbar\bar{\omega}}\tilde{S}^*_{ADCB}, \\
\tilde{S}^*_{CDBA} = e^{\beta\hbar\bar{\omega}}\tilde{S}^*_{ACDB}.
\end{eqnarray}
Combining these frequency relationships it is possible to relate differences of correlation functions to sums as
\begin{eqnarray}
\tilde{S}_{DCBA} -\tilde{S}_{ADCB} &=& \tanh(\beta\hbar\bar{\omega}/2)\left[\tilde{S}_{DCBA} +\tilde{S}_{ADCB}\right],  \\
\tilde{S}^*_{DCBA} -\tilde{S}^*_{ADCB} &=& -\tanh(\beta\hbar\bar{\omega}/2)\left[\tilde{S}^*_{DCBA} +\tilde{S}^*_{ADCB}\right],  \\
\tilde{S}_{CDBA} -\tilde{S}_{ACDB} &=& \tanh(\beta\hbar\bar{\omega}/2)\left[\tilde{S}_{CDBA} +\tilde{S}_{ACDB}\right], \\
\tilde{S}^*_{CDBA} -\tilde{S}^*_{ACDB} &=& -\tanh(\beta\hbar\bar{\omega}/2)\left[\tilde{S}^*_{CDBA} +\tilde{S}^*_{ACDB}\right].
\end{eqnarray}
Which are the relationships that ultimately allow us to write Eq.~(\ref{eq:R3_std}) as Eq.~(\ref{eq:R3_freqq}).

\section{Relation between $\tilde{K}(\omega_1,\omega_2,\omega_3)$ and $\tilde{S}(\omega_1,\omega_2,\omega_3)$}
\label{sec:AppB}

Here we derive the $f(\omega_1,\omega_2,\omega_3)$ conversion factor.
Starting from the Kubo transformed correlation function
\begin{eqnarray}
K(t_1,t_2,t_3) &=& \frac{1}{Z\beta^3}\int^{\beta}_0d\lambda \int^{\lambda}_0d\mu\int^{\mu}_0d\nu \: \nonumber \\ && \times Tr\Big[ e^{-(\beta-\lambda)\hat{H}}\hat{C}(t_2)e^{-(\lambda-\mu)\hat{H}}\hat{D}(t_3) \nonumber \\ && \times e^{-(\mu-\nu)\hat{H}}\hat{B}(t_1)e^{-\nu\hat{H}}\hat{A}(0) \Big],
\end{eqnarray}
then writing it in the energy eigenbasis and Fourier transforming gives
\begin{multline}
\tilde{K}(\omega_1,\omega_2,\omega_3) =\frac{1}{Z\beta^3} \sum_{klmn}e^{-\beta E_k}\Big[\int^{\beta}_0d\lambda \int^{\lambda}_0d\mu\int^{\mu}_0d\nu \\ \times e^{\lambda\hbar\omega_{kl}}   e^{\mu\hbar\omega_{lm}}e^{\nu\hbar\omega_{mn}} \Big] C_{kl}D_{lm}B_{mn}A_{nk} \\ \times \delta(\omega_1-\omega_{mn})\delta(\omega_2-\omega_{kl})\delta(\omega_3-\omega_{lm}).
\end{multline}
Enforcing the delta functions allows us to move the bracketed term outside of the sum giving
\begin{eqnarray}
\tilde{K}(\omega_1,\omega_2,\omega_3) &=
&f(\omega_1,\omega_2,\omega_3) \sum_{klmn}e^{-\beta E_n} C_{kl}D_{lm}B_{mn}A_{nk} \nonumber \\ && \times\delta(\omega_1-\omega_{mn})\delta(\omega_2-\omega_{kl})\delta(\omega_3-\omega_{lm}),
\end{eqnarray}
where
\begin{equation}
f(\omega_1,\omega_2,\omega_3) = \frac{1}{\beta^3}\int^{\beta}_0d\lambda \int^{\lambda}_0d\mu\int^{\mu}_0d\nu e^{\lambda\hbar\omega_{2}}e^{\mu\hbar\omega_3}e^{\nu\hbar\omega_{1}}.
\end{equation}
Renaming the indices gives
\begin{equation}
\tilde{K}(\omega_1,\omega_2,\omega_3) =f(\omega_1,\omega_2,\omega_3)\tilde{S}(\omega_1,\omega_2,\omega_3),
\end{equation}
which is Eq.~(\ref{eq:StoK}). The three iterated integrals in $f(\omega_1,\omega_2,\omega_3)$ can be performed analytically giving the explicit form of the conversion factor as
\begin{eqnarray}
f(\omega_1,\omega_2,\omega_3) &=& \frac{1}{\beta^3\hbar^3}\Bigg[ \frac{e^{\beta\hbar\bar{\omega}}-1}{\omega_1\bar{\omega}(\omega_1+\omega_3)} 
+ \frac{e^{\beta\hbar\omega_2}-1}{\omega_2\omega_3(\omega_1+\omega_3)} 
 \nonumber \\ && - \frac{e^{\beta\hbar(\omega_2+\omega_3)}-1}{\omega_1\omega_3(\omega_2+\omega_3)}
\Bigg].
\label{eq:F_factor_red}
\end{eqnarray}

\section{Method of Devane et. al. using the standard correlation function}
\label{sec:AppC}
Here we will derive a correction factor using the method presented in Ref.~\citenum{DeVane2004} for standard quantum correlation functions. Beginning from
\begin{multline}
    R^{(3)}(\omega_1,\omega_2,\omega_3) = \frac{i}{\hbar^3}\Big[(e^{\beta\hbar\bar\omega}-1)(g(\omega_1,\omega_2,\omega_3)-1)\tilde{S}  \\  - (e^{-\beta\hbar\bar\omega}-1)(g(-\omega_1,-\omega_2,-\omega_3)-1) \tilde{S}^* \Big],
\end{multline}
then separating each term into its even and odd part
\begin{multline}
    R^{(3)}(\omega_1,\omega_2,\omega_3) = \\ \frac{i}{\hbar^3}\Big[(k(\omega_1,\omega_2,\omega_3)-k(-\omega_1,-\omega_2,-\omega_3))\textrm{Re}[\tilde{S}] \\ + (k(\omega_1,\omega_2,\omega_3)+k(-\omega_1,-\omega_2,-\omega_3)) \textrm{Im}[\tilde{S}] \Big],
\end{multline}
where
\begin{equation}
    k(\omega_1,\omega_2,\omega_3) = (e^{\beta\hbar\bar\omega}-1)(g(\omega_1,\omega_2,\omega_3)-1).
\end{equation}
Since the terms $k(\omega_1,\omega_2,\omega_3)-k(-\omega_1,-\omega_2,-\omega_3)$ and $k(\omega_1,\omega_2,\omega_3)+k(-\omega_1,-\omega_2,-\omega_3)$ have a different dependence on $\hbar$ one needs to find an approximate relationship between $\textrm{Re}[\tilde{S}]$ and $\textrm{Im}[\tilde{S}]$ which yields a consistent $\hbar\to0$ limit. One relationship that accomplishes this is\cite{DeVane2004}
\begin{equation}
    l(\omega_1,\omega_2,\omega_3) =-\tanh\left[\frac{\beta\hbar}{4}(\omega_1+3\omega_2+2\omega_3) \right].
\end{equation}
which assures that in the classical limit each term is independent of $\hbar$, giving the response as
\begin{multline}
    R^{(3)}(\omega_1,\omega_2,\omega_3) = \frac{i}{\hbar^3}\Big[(k(\omega_1,\omega_2,\omega_3)-k(-\omega_1,-\omega_2,-\omega_3))  \\  + l(\omega_1,\omega_2,\omega_3)(k(\omega_1,\omega_2,\omega_3)+k(-\omega_1,-\omega_2,-\omega_3)) \Big]\textrm{Re}[\tilde{S}]  \\ = h_{S}(\omega_1,\omega_2,\omega_3)\textrm{Re}[\tilde{S}].
\end{multline}
The classical limit of the $h_{S}(\omega_1,\omega_2,\omega_3)$ correction factor is 
\begin{equation}
    \lim_{\hbar\to0}h_{S}(\omega_1,\omega_2,\omega_3) = \frac{\beta^3}{8}\bar{\omega}(\omega^2_3-\omega^2_2),
\end{equation}
which differs from the classical limit derived using the Kubo transformed correlation functions in the main text by a factor of two. It is reassuring that the two methods yield the same frequency dependence in the classical limit despite the differing forms of the two correction factors.

\bibliography{QCF}

\begin{thebibliography}{69}%
\makeatletter
\providecommand \@ifxundefined [1]{%
 \@ifx{#1\undefined}
}%
\providecommand \@ifnum [1]{%
 \ifnum #1\expandafter \@firstoftwo
 \else \expandafter \@secondoftwo
 \fi
}%
\providecommand \@ifx [1]{%
 \ifx #1\expandafter \@firstoftwo
 \else \expandafter \@secondoftwo
 \fi
}%
\providecommand \natexlab [1]{#1}%
\providecommand \enquote  [1]{``#1''}%
\providecommand \bibnamefont  [1]{#1}%
\providecommand \bibfnamefont [1]{#1}%
\providecommand \citenamefont [1]{#1}%
\providecommand \href@noop [0]{\@secondoftwo}%
\providecommand \href [0]{\begingroup \@sanitize@url \@href}%
\providecommand \@href[1]{\@@startlink{#1}\@@href}%
\providecommand \@@href[1]{\endgroup#1\@@endlink}%
\providecommand \@sanitize@url [0]{\catcode `\\12\catcode `\$12\catcode
  `\&12\catcode `\#12\catcode `\^12\catcode `\_12\catcode `\%12\relax}%
\providecommand \@@startlink[1]{}%
\providecommand \@@endlink[0]{}%
\providecommand \url  [0]{\begingroup\@sanitize@url \@url }%
\providecommand \@url [1]{\endgroup\@href {#1}{\urlprefix }}%
\providecommand \urlprefix  [0]{URL }%
\providecommand \Eprint [0]{\href }%
\providecommand \doibase [0]{http://dx.doi.org/}%
\providecommand \selectlanguage [0]{\@gobble}%
\providecommand \bibinfo  [0]{\@secondoftwo}%
\providecommand \bibfield  [0]{\@secondoftwo}%
\providecommand \translation [1]{[#1]}%
\providecommand \BibitemOpen [0]{}%
\providecommand \bibitemStop [0]{}%
\providecommand \bibitemNoStop [0]{.\EOS\space}%
\providecommand \EOS [0]{\spacefactor3000\relax}%
\providecommand \BibitemShut  [1]{\csname bibitem#1\endcsname}%
\let\auto@bib@innerbib\@empty
\bibitem [{\citenamefont {Hamm}\ and\ \citenamefont {Zanni}(2009)}]{Hamm2009}%
  \BibitemOpen
  \bibfield  {author} {\bibinfo {author} {\bibfnamefont {P.}~\bibnamefont
  {Hamm}}\ and\ \bibinfo {author} {\bibfnamefont {M.}~\bibnamefont {Zanni}},\
  }\href {\doibase 10.1017/cbo9780511675935} {\emph {\bibinfo {title} {Concepts
  and Methods of 2D Infrared Spectroscopy}}}\ (\bibinfo  {publisher} {Cambridge
  University Press},\ \bibinfo {year} {2009})\BibitemShut {NoStop}%
\bibitem [{\citenamefont {la~Cour~Jansen}\ \emph {et~al.}(2019)\citenamefont
  {la~Cour~Jansen}, \citenamefont {Saito}, \citenamefont {Jeon},\ and\
  \citenamefont {Cho}}]{Jansen2019}%
  \BibitemOpen
  \bibfield  {author} {\bibinfo {author} {\bibfnamefont {T.}~\bibnamefont
  {la~Cour~Jansen}}, \bibinfo {author} {\bibfnamefont {S.}~\bibnamefont
  {Saito}}, \bibinfo {author} {\bibfnamefont {J.}~\bibnamefont {Jeon}}, \ and\
  \bibinfo {author} {\bibfnamefont {M.}~\bibnamefont {Cho}},\ }\href {\doibase
  10.1063/1.5083966} {\bibfield  {journal} {\bibinfo  {journal} {The Journal of
  Chemical Physics}\ }\textbf {\bibinfo {volume} {150}},\ \bibinfo {pages}
  {100901} (\bibinfo {year} {2019})}\BibitemShut {NoStop}%
\bibitem [{\citenamefont {Biswas}\ \emph {et~al.}(2022)\citenamefont {Biswas},
  \citenamefont {Kim}, \citenamefont {Zhang},\ and\ \citenamefont
  {Scholes}}]{Biswas2022}%
  \BibitemOpen
  \bibfield  {author} {\bibinfo {author} {\bibfnamefont {S.}~\bibnamefont
  {Biswas}}, \bibinfo {author} {\bibfnamefont {J.}~\bibnamefont {Kim}},
  \bibinfo {author} {\bibfnamefont {X.}~\bibnamefont {Zhang}}, \ and\ \bibinfo
  {author} {\bibfnamefont {G.~D.}\ \bibnamefont {Scholes}},\ }\href {\doibase
  10.1021/acs.chemrev.1c00623} {\bibfield  {journal} {\bibinfo  {journal}
  {Chemical Reviews}\ }\textbf {\bibinfo {volume} {122}},\ \bibinfo {pages}
  {4257} (\bibinfo {year} {2022})}\BibitemShut {NoStop}%
\bibitem [{\citenamefont {Hamm}, \citenamefont {Lim},\ and\ \citenamefont
  {Hochstrasser}(1998)}]{Hamm1998}%
  \BibitemOpen
  \bibfield  {author} {\bibinfo {author} {\bibfnamefont {P.}~\bibnamefont
  {Hamm}}, \bibinfo {author} {\bibfnamefont {M.}~\bibnamefont {Lim}}, \ and\
  \bibinfo {author} {\bibfnamefont {R.~M.}\ \bibnamefont {Hochstrasser}},\
  }\href {\doibase 10.1021/jp9813286} {\bibfield  {journal} {\bibinfo
  {journal} {The Journal of Physical Chemistry B}\ }\textbf {\bibinfo {volume}
  {102}},\ \bibinfo {pages} {6123} (\bibinfo {year} {1998})}\BibitemShut
  {NoStop}%
\bibitem [{\citenamefont {Zanni}\ and\ \citenamefont
  {Hochstrasser}(2001)}]{Zanni2001}%
  \BibitemOpen
  \bibfield  {author} {\bibinfo {author} {\bibfnamefont {M.~T.}\ \bibnamefont
  {Zanni}}\ and\ \bibinfo {author} {\bibfnamefont {R.~M.}\ \bibnamefont
  {Hochstrasser}},\ }\href {\doibase 10.1016/s0959-440x(00)00243-8} {\bibfield
  {journal} {\bibinfo  {journal} {Current Opinion in Structural Biology}\
  }\textbf {\bibinfo {volume} {11}},\ \bibinfo {pages} {516} (\bibinfo {year}
  {2001})}\BibitemShut {NoStop}%
\bibitem [{\citenamefont {Savolainen}, \citenamefont {Ahmed},\ and\
  \citenamefont {Hamm}(2013)}]{Savolainen2013}%
  \BibitemOpen
  \bibfield  {author} {\bibinfo {author} {\bibfnamefont {J.}~\bibnamefont
  {Savolainen}}, \bibinfo {author} {\bibfnamefont {S.}~\bibnamefont {Ahmed}}, \
  and\ \bibinfo {author} {\bibfnamefont {P.}~\bibnamefont {Hamm}},\ }\href
  {\doibase 10.1073/pnas.1317459110} {\bibfield  {journal} {\bibinfo  {journal}
  {Proceedings of the National Academy of Sciences}\ }\textbf {\bibinfo
  {volume} {110}},\ \bibinfo {pages} {20402} (\bibinfo {year}
  {2013})}\BibitemShut {NoStop}%
\bibitem [{\citenamefont {Cho}(1999)}]{Cho1999}%
  \BibitemOpen
  \bibfield  {author} {\bibinfo {author} {\bibfnamefont {M.}~\bibnamefont
  {Cho}},\ }\href {\doibase 10.1063/1.479711} {\bibfield  {journal} {\bibinfo
  {journal} {The Journal of Chemical Physics}\ }\textbf {\bibinfo {volume}
  {111}},\ \bibinfo {pages} {4140} (\bibinfo {year} {1999})},\ \Eprint
  {http://arxiv.org/abs/https://doi.org/10.1063/1.479711}
  {https://doi.org/10.1063/1.479711} \BibitemShut {NoStop}%
\bibitem [{\citenamefont {Xiong}\ \emph {et~al.}(2011)\citenamefont {Xiong},
  \citenamefont {Laaser}, \citenamefont {Mehlenbacher},\ and\ \citenamefont
  {Zanni}}]{Xiong2011}%
  \BibitemOpen
  \bibfield  {author} {\bibinfo {author} {\bibfnamefont {W.}~\bibnamefont
  {Xiong}}, \bibinfo {author} {\bibfnamefont {J.~E.}\ \bibnamefont {Laaser}},
  \bibinfo {author} {\bibfnamefont {R.~D.}\ \bibnamefont {Mehlenbacher}}, \
  and\ \bibinfo {author} {\bibfnamefont {M.~T.}\ \bibnamefont {Zanni}},\ }\href
  {\doibase 10.1073/pnas.1115055108} {\bibfield  {journal} {\bibinfo  {journal}
  {Proceedings of the National Academy of Sciences}\ }\textbf {\bibinfo
  {volume} {108}},\ \bibinfo {pages} {20902} (\bibinfo {year}
  {2011})}\BibitemShut {NoStop}%
\bibitem [{\citenamefont {Berne}, \citenamefont {Jortner},\ and\ \citenamefont
  {Gordon}(1967)}]{Berne1967}%
  \BibitemOpen
  \bibfield  {author} {\bibinfo {author} {\bibfnamefont {B.~J.}\ \bibnamefont
  {Berne}}, \bibinfo {author} {\bibfnamefont {J.}~\bibnamefont {Jortner}}, \
  and\ \bibinfo {author} {\bibfnamefont {R.}~\bibnamefont {Gordon}},\ }\href
  {\doibase 10.1063/1.1712140} {\bibfield  {journal} {\bibinfo  {journal} {The
  Journal of Chemical Physics}\ }\textbf {\bibinfo {volume} {47}},\ \bibinfo
  {pages} {1600} (\bibinfo {year} {1967})}\BibitemShut {NoStop}%
\bibitem [{\citenamefont {Mukamel}(1985)}]{Mukamel1985}%
  \BibitemOpen
  \bibfield  {author} {\bibinfo {author} {\bibfnamefont {S.}~\bibnamefont
  {Mukamel}},\ }\href {\doibase 10.1021/j100253a008} {\bibfield  {journal}
  {\bibinfo  {journal} {The Journal of Physical Chemistry}\ }\textbf {\bibinfo
  {volume} {89}},\ \bibinfo {pages} {1077} (\bibinfo {year}
  {1985})}\BibitemShut {NoStop}%
\bibitem [{\citenamefont {Ahlborn}, \citenamefont {Space},\ and\ \citenamefont
  {Moore}(2000)}]{Ahlborn2000}%
  \BibitemOpen
  \bibfield  {author} {\bibinfo {author} {\bibfnamefont {H.}~\bibnamefont
  {Ahlborn}}, \bibinfo {author} {\bibfnamefont {B.}~\bibnamefont {Space}}, \
  and\ \bibinfo {author} {\bibfnamefont {P.~B.}\ \bibnamefont {Moore}},\ }\href
  {\doibase 10.1063/1.481408} {\bibfield  {journal} {\bibinfo  {journal} {The
  Journal of Chemical Physics}\ }\textbf {\bibinfo {volume} {112}},\ \bibinfo
  {pages} {8083} (\bibinfo {year} {2000})}\BibitemShut {NoStop}%
\bibitem [{\citenamefont {Lawrence}\ and\ \citenamefont
  {Skinner}(2002)}]{Lawrence2002}%
  \BibitemOpen
  \bibfield  {author} {\bibinfo {author} {\bibfnamefont {C.~P.}\ \bibnamefont
  {Lawrence}}\ and\ \bibinfo {author} {\bibfnamefont {J.~L.}\ \bibnamefont
  {Skinner}},\ }\href {\doibase 10.1063/1.1514652} {\bibfield  {journal}
  {\bibinfo  {journal} {The Journal of Chemical Physics}\ }\textbf {\bibinfo
  {volume} {117}},\ \bibinfo {pages} {8847} (\bibinfo {year}
  {2002})}\BibitemShut {NoStop}%
\bibitem [{\citenamefont {McRobbie}\ and\ \citenamefont
  {Geva}(2009)}]{McRobbie2009}%
  \BibitemOpen
  \bibfield  {author} {\bibinfo {author} {\bibfnamefont {P.~L.}\ \bibnamefont
  {McRobbie}}\ and\ \bibinfo {author} {\bibfnamefont {E.}~\bibnamefont
  {Geva}},\ }\href {\doibase 10.1021/jp905305t} {\bibfield  {journal} {\bibinfo
   {journal} {The Journal of Physical Chemistry A}\ }\textbf {\bibinfo {volume}
  {113}},\ \bibinfo {pages} {10425} (\bibinfo {year} {2009})}\BibitemShut
  {NoStop}%
\bibitem [{\citenamefont {Liu}\ \emph {et~al.}(2009)\citenamefont {Liu},
  \citenamefont {Miller}, \citenamefont {Paesani}, \citenamefont {Zhang},\ and\
  \citenamefont {Case}}]{Liu2009}%
  \BibitemOpen
  \bibfield  {author} {\bibinfo {author} {\bibfnamefont {J.}~\bibnamefont
  {Liu}}, \bibinfo {author} {\bibfnamefont {W.~H.}\ \bibnamefont {Miller}},
  \bibinfo {author} {\bibfnamefont {F.}~\bibnamefont {Paesani}}, \bibinfo
  {author} {\bibfnamefont {W.}~\bibnamefont {Zhang}}, \ and\ \bibinfo {author}
  {\bibfnamefont {D.~A.}\ \bibnamefont {Case}},\ }\href {\doibase
  10.1063/1.3254372} {\bibfield  {journal} {\bibinfo  {journal} {The Journal of
  Chemical Physics}\ }\textbf {\bibinfo {volume} {131}},\ \bibinfo {pages}
  {164509} (\bibinfo {year} {2009})}\BibitemShut {NoStop}%
\bibitem [{\citenamefont {Witt}\ \emph {et~al.}(2009)\citenamefont {Witt},
  \citenamefont {Ivanov}, \citenamefont {Shiga}, \citenamefont {Forbert},\ and\
  \citenamefont {Marx}}]{Witt2009}%
  \BibitemOpen
  \bibfield  {author} {\bibinfo {author} {\bibfnamefont {A.}~\bibnamefont
  {Witt}}, \bibinfo {author} {\bibfnamefont {S.~D.}\ \bibnamefont {Ivanov}},
  \bibinfo {author} {\bibfnamefont {M.}~\bibnamefont {Shiga}}, \bibinfo
  {author} {\bibfnamefont {H.}~\bibnamefont {Forbert}}, \ and\ \bibinfo
  {author} {\bibfnamefont {D.}~\bibnamefont {Marx}},\ }\href {\doibase
  10.1063/1.3125009} {\bibfield  {journal} {\bibinfo  {journal} {The Journal of
  Chemical Physics}\ }\textbf {\bibinfo {volume} {130}},\ \bibinfo {pages}
  {194510} (\bibinfo {year} {2009})},\ \Eprint
  {http://arxiv.org/abs/https://doi.org/10.1063/1.3125009}
  {https://doi.org/10.1063/1.3125009} \BibitemShut {NoStop}%
\bibitem [{\citenamefont {Liu}\ and\ \citenamefont {Zhang}(2016)}]{Liu2016}%
  \BibitemOpen
  \bibfield  {author} {\bibinfo {author} {\bibfnamefont {J.}~\bibnamefont
  {Liu}}\ and\ \bibinfo {author} {\bibfnamefont {Z.}~\bibnamefont {Zhang}},\
  }\href {\doibase 10.1063/1.4939953} {\bibfield  {journal} {\bibinfo
  {journal} {The Journal of Chemical Physics}\ }\textbf {\bibinfo {volume}
  {144}},\ \bibinfo {pages} {034307} (\bibinfo {year} {2016})},\ \Eprint
  {http://arxiv.org/abs/https://doi.org/10.1063/1.4939953}
  {https://doi.org/10.1063/1.4939953} \BibitemShut {NoStop}%
\bibitem [{\citenamefont {Willatt}, \citenamefont {Ceriotti},\ and\
  \citenamefont {Althorpe}(2018)}]{Willat2018}%
  \BibitemOpen
  \bibfield  {author} {\bibinfo {author} {\bibfnamefont {M.~J.}\ \bibnamefont
  {Willatt}}, \bibinfo {author} {\bibfnamefont {M.}~\bibnamefont {Ceriotti}}, \
  and\ \bibinfo {author} {\bibfnamefont {S.~C.}\ \bibnamefont {Althorpe}},\
  }\href {\doibase 10.1063/1.5004808} {\bibfield  {journal} {\bibinfo
  {journal} {The Journal of Chemical Physics}\ }\textbf {\bibinfo {volume}
  {148}},\ \bibinfo {pages} {102336} (\bibinfo {year} {2018})},\ \Eprint
  {http://arxiv.org/abs/https://doi.org/10.1063/1.5004808}
  {https://doi.org/10.1063/1.5004808} \BibitemShut {NoStop}%
\bibitem [{\citenamefont {Trenins}, \citenamefont {Willatt},\ and\
  \citenamefont {Althorpe}(2019)}]{Trenins2019}%
  \BibitemOpen
  \bibfield  {author} {\bibinfo {author} {\bibfnamefont {G.}~\bibnamefont
  {Trenins}}, \bibinfo {author} {\bibfnamefont {M.~J.}\ \bibnamefont
  {Willatt}}, \ and\ \bibinfo {author} {\bibfnamefont {S.~C.}\ \bibnamefont
  {Althorpe}},\ }\href {\doibase 10.1063/1.5100587} {\bibfield  {journal}
  {\bibinfo  {journal} {The Journal of Chemical Physics}\ }\textbf {\bibinfo
  {volume} {151}},\ \bibinfo {pages} {054109} (\bibinfo {year}
  {2019})}\BibitemShut {NoStop}%
\bibitem [{\citenamefont {Benson}\ and\ \citenamefont
  {Althorpe}(2021)}]{Benson2021}%
  \BibitemOpen
  \bibfield  {author} {\bibinfo {author} {\bibfnamefont {R.~L.}\ \bibnamefont
  {Benson}}\ and\ \bibinfo {author} {\bibfnamefont {S.~C.}\ \bibnamefont
  {Althorpe}},\ }\href {\doibase 10.1063/5.0056829} {\bibfield  {journal}
  {\bibinfo  {journal} {The Journal of Chemical Physics}\ }\textbf {\bibinfo
  {volume} {155}},\ \bibinfo {pages} {104107} (\bibinfo {year}
  {2021})}\BibitemShut {NoStop}%
\bibitem [{\citenamefont {Mukamel}(1995)}]{Mukamel_Book}%
  \BibitemOpen
  \bibfield  {author} {\bibinfo {author} {\bibfnamefont {S.}~\bibnamefont
  {Mukamel}},\ }\href
  {https://www.amazon.com/Principles-Nonlinear-Optical-Spectroscopy-Mukamel/dp/0195092783?SubscriptionId=0JYN1NVW651KCA56C102&tag=techkie-20&linkCode=xm2&camp=2025&creative=165953&creativeASIN=0195092783}
  {\emph {\bibinfo {title} {Principles of Nonlinear Optical Spectroscopy}}}\
  (\bibinfo  {publisher} {Oxford University Press},\ \bibinfo {year}
  {1995})\BibitemShut {NoStop}%
\bibitem [{\citenamefont {Cho}(2009)}]{Cho_Book}%
  \BibitemOpen
  \bibfield  {author} {\bibinfo {author} {\bibfnamefont {M.}~\bibnamefont
  {Cho}},\ }\href
  {https://www.amazon.com/Two-Dimensional-Optical-Spectroscopy-Minhaeng-Cho/dp/1420084291?SubscriptionId=0JYN1NVW651KCA56C102&tag=techkie-20&linkCode=xm2&camp=2025&creative=165953&creativeASIN=1420084291}
  {\emph {\bibinfo {title} {Two-Dimensional Optical Spectroscopy}}}\ (\bibinfo
  {publisher} {CRC Press},\ \bibinfo {year} {2009})\BibitemShut {NoStop}%
\bibitem [{\citenamefont {Lee}\ and\ \citenamefont
  {Tuckerman}(2007)}]{Lee2007}%
  \BibitemOpen
  \bibfield  {author} {\bibinfo {author} {\bibfnamefont {H.-S.}\ \bibnamefont
  {Lee}}\ and\ \bibinfo {author} {\bibfnamefont {M.~E.}\ \bibnamefont
  {Tuckerman}},\ }\href {\doibase 10.1063/1.2718521} {\bibfield  {journal}
  {\bibinfo  {journal} {The Journal of Chemical Physics}\ }\textbf {\bibinfo
  {volume} {126}},\ \bibinfo {pages} {164501} (\bibinfo {year}
  {2007})}\BibitemShut {NoStop}%
\bibitem [{\citenamefont {Ceriotti}\ and\ \citenamefont
  {Manolopoulos}(2012)}]{Ceriotti2012}%
  \BibitemOpen
  \bibfield  {author} {\bibinfo {author} {\bibfnamefont {M.}~\bibnamefont
  {Ceriotti}}\ and\ \bibinfo {author} {\bibfnamefont {D.~E.}\ \bibnamefont
  {Manolopoulos}},\ }\href {\doibase 10.1103/physrevlett.109.100604} {\bibfield
   {journal} {\bibinfo  {journal} {Physical Review Letters}\ }\textbf {\bibinfo
  {volume} {109}} (\bibinfo {year} {2012}),\
  10.1103/physrevlett.109.100604}\BibitemShut {NoStop}%
\bibitem [{\citenamefont {Marsalek}\ and\ \citenamefont
  {Markland}(2016)}]{Marsalek2016}%
  \BibitemOpen
  \bibfield  {author} {\bibinfo {author} {\bibfnamefont {O.}~\bibnamefont
  {Marsalek}}\ and\ \bibinfo {author} {\bibfnamefont {T.~E.}\ \bibnamefont
  {Markland}},\ }\href {\doibase 10.1063/1.4941093} {\bibfield  {journal}
  {\bibinfo  {journal} {The Journal of Chemical Physics}\ }\textbf {\bibinfo
  {volume} {144}},\ \bibinfo {pages} {054112} (\bibinfo {year}
  {2016})}\BibitemShut {NoStop}%
\bibitem [{\citenamefont {Marsalek}\ and\ \citenamefont
  {Markland}(2017)}]{Marsalek2017}%
  \BibitemOpen
  \bibfield  {author} {\bibinfo {author} {\bibfnamefont {O.}~\bibnamefont
  {Marsalek}}\ and\ \bibinfo {author} {\bibfnamefont {T.~E.}\ \bibnamefont
  {Markland}},\ }\href {\doibase 10.1021/acs.jpclett.7b00391} {\bibfield
  {journal} {\bibinfo  {journal} {The Journal of Physical Chemistry Letters}\
  }\textbf {\bibinfo {volume} {8}},\ \bibinfo {pages} {1545} (\bibinfo {year}
  {2017})}\BibitemShut {NoStop}%
\bibitem [{\citenamefont {Sharma}\ \emph {et~al.}(2021)\citenamefont {Sharma},
  \citenamefont {Tran}, \citenamefont {Pongratz}, \citenamefont {Galazzo},
  \citenamefont {Zhurko}, \citenamefont {Bordignon}, \citenamefont {Kast},
  \citenamefont {Neese},\ and\ \citenamefont {Marx}}]{Sharma2021}%
  \BibitemOpen
  \bibfield  {author} {\bibinfo {author} {\bibfnamefont {B.}~\bibnamefont
  {Sharma}}, \bibinfo {author} {\bibfnamefont {V.~A.}\ \bibnamefont {Tran}},
  \bibinfo {author} {\bibfnamefont {T.}~\bibnamefont {Pongratz}}, \bibinfo
  {author} {\bibfnamefont {L.}~\bibnamefont {Galazzo}}, \bibinfo {author}
  {\bibfnamefont {I.}~\bibnamefont {Zhurko}}, \bibinfo {author} {\bibfnamefont
  {E.}~\bibnamefont {Bordignon}}, \bibinfo {author} {\bibfnamefont {S.~M.}\
  \bibnamefont {Kast}}, \bibinfo {author} {\bibfnamefont {F.}~\bibnamefont
  {Neese}}, \ and\ \bibinfo {author} {\bibfnamefont {D.}~\bibnamefont {Marx}},\
  }\href {\doibase 10.1021/acs.jctc.1c00582} {\bibfield  {journal} {\bibinfo
  {journal} {Journal of Chemical Theory and Computation}\ }\textbf {\bibinfo
  {volume} {17}},\ \bibinfo {pages} {6366} (\bibinfo {year}
  {2021})}\BibitemShut {NoStop}%
\bibitem [{\citenamefont {Behler}\ and\ \citenamefont
  {Parrinello}(2007)}]{Behler2007}%
  \BibitemOpen
  \bibfield  {author} {\bibinfo {author} {\bibfnamefont {J.}~\bibnamefont
  {Behler}}\ and\ \bibinfo {author} {\bibfnamefont {M.}~\bibnamefont
  {Parrinello}},\ }\href {\doibase 10.1103/physrevlett.98.146401} {\bibfield
  {journal} {\bibinfo  {journal} {Physical Review Letters}\ }\textbf {\bibinfo
  {volume} {98}} (\bibinfo {year} {2007}),\
  10.1103/physrevlett.98.146401}\BibitemShut {NoStop}%
\bibitem [{\citenamefont {Grisafi}\ \emph {et~al.}(2018)\citenamefont
  {Grisafi}, \citenamefont {Wilkins}, \citenamefont {Cs{\'{a}}nyi},\ and\
  \citenamefont {Ceriotti}}]{Grisafi2018}%
  \BibitemOpen
  \bibfield  {author} {\bibinfo {author} {\bibfnamefont {A.}~\bibnamefont
  {Grisafi}}, \bibinfo {author} {\bibfnamefont {D.~M.}\ \bibnamefont
  {Wilkins}}, \bibinfo {author} {\bibfnamefont {G.}~\bibnamefont
  {Cs{\'{a}}nyi}}, \ and\ \bibinfo {author} {\bibfnamefont {M.}~\bibnamefont
  {Ceriotti}},\ }\href {\doibase 10.1103/physrevlett.120.036002} {\bibfield
  {journal} {\bibinfo  {journal} {Physical Review Letters}\ }\textbf {\bibinfo
  {volume} {120}} (\bibinfo {year} {2018}),\
  10.1103/physrevlett.120.036002}\BibitemShut {NoStop}%
\bibitem [{\citenamefont {Zuo}\ \emph {et~al.}(2020)\citenamefont {Zuo},
  \citenamefont {Chen}, \citenamefont {Li}, \citenamefont {Deng}, \citenamefont
  {Chen}, \citenamefont {Behler}, \citenamefont {Cs{\'{a}}nyi}, \citenamefont
  {Shapeev}, \citenamefont {Thompson}, \citenamefont {Wood},\ and\
  \citenamefont {Ong}}]{Zuo2020}%
  \BibitemOpen
  \bibfield  {author} {\bibinfo {author} {\bibfnamefont {Y.}~\bibnamefont
  {Zuo}}, \bibinfo {author} {\bibfnamefont {C.}~\bibnamefont {Chen}}, \bibinfo
  {author} {\bibfnamefont {X.}~\bibnamefont {Li}}, \bibinfo {author}
  {\bibfnamefont {Z.}~\bibnamefont {Deng}}, \bibinfo {author} {\bibfnamefont
  {Y.}~\bibnamefont {Chen}}, \bibinfo {author} {\bibfnamefont {J.}~\bibnamefont
  {Behler}}, \bibinfo {author} {\bibfnamefont {G.}~\bibnamefont
  {Cs{\'{a}}nyi}}, \bibinfo {author} {\bibfnamefont {A.~V.}\ \bibnamefont
  {Shapeev}}, \bibinfo {author} {\bibfnamefont {A.~P.}\ \bibnamefont
  {Thompson}}, \bibinfo {author} {\bibfnamefont {M.~A.}\ \bibnamefont {Wood}},
  \ and\ \bibinfo {author} {\bibfnamefont {S.~P.}\ \bibnamefont {Ong}},\ }\href
  {\doibase 10.1021/acs.jpca.9b08723} {\bibfield  {journal} {\bibinfo
  {journal} {The Journal of Physical Chemistry A}\ }\textbf {\bibinfo {volume}
  {124}},\ \bibinfo {pages} {731} (\bibinfo {year} {2020})}\BibitemShut
  {NoStop}%
\bibitem [{\citenamefont {Chen}\ \emph {et~al.}(2020)\citenamefont {Chen},
  \citenamefont {Zuehlsdorff}, \citenamefont {Morawietz}, \citenamefont
  {Isborn},\ and\ \citenamefont {Markland}}]{Chen2020}%
  \BibitemOpen
  \bibfield  {author} {\bibinfo {author} {\bibfnamefont {M.~S.}\ \bibnamefont
  {Chen}}, \bibinfo {author} {\bibfnamefont {T.~J.}\ \bibnamefont
  {Zuehlsdorff}}, \bibinfo {author} {\bibfnamefont {T.}~\bibnamefont
  {Morawietz}}, \bibinfo {author} {\bibfnamefont {C.~M.}\ \bibnamefont
  {Isborn}}, \ and\ \bibinfo {author} {\bibfnamefont {T.~E.}\ \bibnamefont
  {Markland}},\ }\href {\doibase 10.1021/acs.jpclett.0c02168} {\bibfield
  {journal} {\bibinfo  {journal} {The Journal of Physical Chemistry Letters}\
  }\textbf {\bibinfo {volume} {11}},\ \bibinfo {pages} {7559} (\bibinfo {year}
  {2020})}\BibitemShut {NoStop}%
\bibitem [{\citenamefont {Chen}\ \emph {et~al.}(2021)\citenamefont {Chen},
  \citenamefont {Morawietz}, \citenamefont {Mori}, \citenamefont {Markland},\
  and\ \citenamefont {Artrith}}]{Chen2021}%
  \BibitemOpen
  \bibfield  {author} {\bibinfo {author} {\bibfnamefont {M.~S.}\ \bibnamefont
  {Chen}}, \bibinfo {author} {\bibfnamefont {T.}~\bibnamefont {Morawietz}},
  \bibinfo {author} {\bibfnamefont {H.}~\bibnamefont {Mori}}, \bibinfo {author}
  {\bibfnamefont {T.~E.}\ \bibnamefont {Markland}}, \ and\ \bibinfo {author}
  {\bibfnamefont {N.}~\bibnamefont {Artrith}},\ }\href {\doibase
  10.1063/5.0063880} {\bibfield  {journal} {\bibinfo  {journal} {The Journal of
  Chemical Physics}\ }\textbf {\bibinfo {volume} {155}},\ \bibinfo {pages}
  {074801} (\bibinfo {year} {2021})}\BibitemShut {NoStop}%
\bibitem [{\citenamefont {Mukamel}, \citenamefont {Khidekel},\ and\
  \citenamefont {Chernyak}(1996)}]{Mukamel1996}%
  \BibitemOpen
  \bibfield  {author} {\bibinfo {author} {\bibfnamefont {S.}~\bibnamefont
  {Mukamel}}, \bibinfo {author} {\bibfnamefont {V.}~\bibnamefont {Khidekel}}, \
  and\ \bibinfo {author} {\bibfnamefont {V.}~\bibnamefont {Chernyak}},\ }\href
  {\doibase 10.1103/physreve.53.r1} {\bibfield  {journal} {\bibinfo  {journal}
  {Physical Review E}\ }\textbf {\bibinfo {volume} {53}},\ \bibinfo {pages}
  {R1} (\bibinfo {year} {1996})}\BibitemShut {NoStop}%
\bibitem [{\citenamefont {Kryvohuz}\ and\ \citenamefont
  {Cao}(2005)}]{Kryvohuz2005}%
  \BibitemOpen
  \bibfield  {author} {\bibinfo {author} {\bibfnamefont {M.}~\bibnamefont
  {Kryvohuz}}\ and\ \bibinfo {author} {\bibfnamefont {J.}~\bibnamefont {Cao}},\
  }\href {\doibase 10.1103/physrevlett.95.180405} {\bibfield  {journal}
  {\bibinfo  {journal} {Physical Review Letters}\ }\textbf {\bibinfo {volume}
  {95}} (\bibinfo {year} {2005}),\ 10.1103/physrevlett.95.180405}\BibitemShut
  {NoStop}%
\bibitem [{\citenamefont {l.~C.~Jansen}, \citenamefont {Snijders},\ and\
  \citenamefont {Duppen}(2000)}]{Jansen2000}%
  \BibitemOpen
  \bibfield  {author} {\bibinfo {author} {\bibfnamefont {T.}~\bibnamefont
  {l.~C.~Jansen}}, \bibinfo {author} {\bibfnamefont {J.~G.}\ \bibnamefont
  {Snijders}}, \ and\ \bibinfo {author} {\bibfnamefont {K.}~\bibnamefont
  {Duppen}},\ }\href {\doibase 10.1063/1.481795} {\bibfield  {journal}
  {\bibinfo  {journal} {The Journal of Chemical Physics}\ }\textbf {\bibinfo
  {volume} {113}},\ \bibinfo {pages} {307} (\bibinfo {year} {2000})},\ \Eprint
  {http://arxiv.org/abs/https://doi.org/10.1063/1.481795}
  {https://doi.org/10.1063/1.481795} \BibitemShut {NoStop}%
\bibitem [{\citenamefont {l.~C.~Jansen}, \citenamefont {Duppen},\ and\
  \citenamefont {Snijders}(2003)}]{Jansen2003}%
  \BibitemOpen
  \bibfield  {author} {\bibinfo {author} {\bibfnamefont {T.}~\bibnamefont
  {l.~C.~Jansen}}, \bibinfo {author} {\bibfnamefont {K.}~\bibnamefont
  {Duppen}}, \ and\ \bibinfo {author} {\bibfnamefont {J.~G.}\ \bibnamefont
  {Snijders}},\ }\href {\doibase 10.1103/physrevb.67.134206} {\bibfield
  {journal} {\bibinfo  {journal} {Physical Review B}\ }\textbf {\bibinfo
  {volume} {67}} (\bibinfo {year} {2003}),\
  10.1103/physrevb.67.134206}\BibitemShut {NoStop}%
\bibitem [{\citenamefont {Yagasaki}\ and\ \citenamefont
  {Saito}(2008)}]{Yagasaki2008}%
  \BibitemOpen
  \bibfield  {author} {\bibinfo {author} {\bibfnamefont {T.}~\bibnamefont
  {Yagasaki}}\ and\ \bibinfo {author} {\bibfnamefont {S.}~\bibnamefont
  {Saito}},\ }\href {\doibase 10.1063/1.2903470} {\bibfield  {journal}
  {\bibinfo  {journal} {The Journal of Chemical Physics}\ }\textbf {\bibinfo
  {volume} {128}},\ \bibinfo {pages} {154521} (\bibinfo {year}
  {2008})}\BibitemShut {NoStop}%
\bibitem [{\citenamefont {Hasegawa}\ and\ \citenamefont
  {Tanimura}(2006)}]{Hasegawa2006}%
  \BibitemOpen
  \bibfield  {author} {\bibinfo {author} {\bibfnamefont {T.}~\bibnamefont
  {Hasegawa}}\ and\ \bibinfo {author} {\bibfnamefont {Y.}~\bibnamefont
  {Tanimura}},\ }\href {\doibase 10.1063/1.2217947} {\bibfield  {journal}
  {\bibinfo  {journal} {The Journal of Chemical Physics}\ }\textbf {\bibinfo
  {volume} {125}},\ \bibinfo {pages} {074512} (\bibinfo {year}
  {2006})}\BibitemShut {NoStop}%
\bibitem [{\citenamefont {Yagasaki}\ and\ \citenamefont
  {Saito}(2009)}]{Yagasaki2009}%
  \BibitemOpen
  \bibfield  {author} {\bibinfo {author} {\bibfnamefont {T.}~\bibnamefont
  {Yagasaki}}\ and\ \bibinfo {author} {\bibfnamefont {S.}~\bibnamefont
  {Saito}},\ }\href {\doibase 10.1021/ar900007s} {\bibfield  {journal}
  {\bibinfo  {journal} {Accounts of Chemical Research}\ }\textbf {\bibinfo
  {volume} {42}},\ \bibinfo {pages} {1250} (\bibinfo {year}
  {2009})}\BibitemShut {NoStop}%
\bibitem [{\citenamefont {Begu{\v{s}}i{\'{c}}}\ \emph
  {et~al.}(2022)\citenamefont {Begu{\v{s}}i{\'{c}}}, \citenamefont {Tao},
  \citenamefont {Blake},\ and\ \citenamefont {Miller}}]{Begui2022}%
  \BibitemOpen
  \bibfield  {author} {\bibinfo {author} {\bibfnamefont {T.}~\bibnamefont
  {Begu{\v{s}}i{\'{c}}}}, \bibinfo {author} {\bibfnamefont {X.}~\bibnamefont
  {Tao}}, \bibinfo {author} {\bibfnamefont {G.~A.}\ \bibnamefont {Blake}}, \
  and\ \bibinfo {author} {\bibfnamefont {T.~F.}\ \bibnamefont {Miller}},\
  }\href {\doibase 10.1063/5.0087156} {\bibfield  {journal} {\bibinfo
  {journal} {The Journal of Chemical Physics}\ }\textbf {\bibinfo {volume}
  {156}},\ \bibinfo {pages} {131102} (\bibinfo {year} {2022})}\BibitemShut
  {NoStop}%
\bibitem [{\citenamefont {DeVane}\ \emph {et~al.}(2003)\citenamefont {DeVane},
  \citenamefont {Ridley}, \citenamefont {Space},\ and\ \citenamefont
  {Keyes}}]{DeVane2003}%
  \BibitemOpen
  \bibfield  {author} {\bibinfo {author} {\bibfnamefont {R.}~\bibnamefont
  {DeVane}}, \bibinfo {author} {\bibfnamefont {C.}~\bibnamefont {Ridley}},
  \bibinfo {author} {\bibfnamefont {B.}~\bibnamefont {Space}}, \ and\ \bibinfo
  {author} {\bibfnamefont {T.}~\bibnamefont {Keyes}},\ }\href {\doibase
  10.1063/1.1601607} {\bibfield  {journal} {\bibinfo  {journal} {The Journal of
  Chemical Physics}\ }\textbf {\bibinfo {volume} {119}},\ \bibinfo {pages}
  {6073} (\bibinfo {year} {2003})}\BibitemShut {NoStop}%
\bibitem [{\citenamefont {DeVane}\ \emph {et~al.}(2004)\citenamefont {DeVane},
  \citenamefont {Space}, \citenamefont {Perry}, \citenamefont {Neipert},
  \citenamefont {Ridley},\ and\ \citenamefont {Keyes}}]{DeVane2004}%
  \BibitemOpen
  \bibfield  {author} {\bibinfo {author} {\bibfnamefont {R.}~\bibnamefont
  {DeVane}}, \bibinfo {author} {\bibfnamefont {B.}~\bibnamefont {Space}},
  \bibinfo {author} {\bibfnamefont {A.}~\bibnamefont {Perry}}, \bibinfo
  {author} {\bibfnamefont {C.}~\bibnamefont {Neipert}}, \bibinfo {author}
  {\bibfnamefont {C.}~\bibnamefont {Ridley}}, \ and\ \bibinfo {author}
  {\bibfnamefont {T.}~\bibnamefont {Keyes}},\ }\href {\doibase
  10.1063/1.1776119} {\bibfield  {journal} {\bibinfo  {journal} {The Journal of
  Chemical Physics}\ }\textbf {\bibinfo {volume} {121}},\ \bibinfo {pages}
  {3688} (\bibinfo {year} {2004})}\BibitemShut {NoStop}%
\bibitem [{\citenamefont {DeVane}\ \emph {et~al.}(2005)\citenamefont {DeVane},
  \citenamefont {Ridley}, \citenamefont {Space},\ and\ \citenamefont
  {Keyes}}]{DeVane2005}%
  \BibitemOpen
  \bibfield  {author} {\bibinfo {author} {\bibfnamefont {R.}~\bibnamefont
  {DeVane}}, \bibinfo {author} {\bibfnamefont {C.}~\bibnamefont {Ridley}},
  \bibinfo {author} {\bibfnamefont {B.}~\bibnamefont {Space}}, \ and\ \bibinfo
  {author} {\bibfnamefont {T.}~\bibnamefont {Keyes}},\ }\href {\doibase
  10.1063/1.2038768} {\bibfield  {journal} {\bibinfo  {journal} {The Journal of
  Chemical Physics}\ }\textbf {\bibinfo {volume} {123}},\ \bibinfo {pages}
  {194507} (\bibinfo {year} {2005})}\BibitemShut {NoStop}%
\bibitem [{\citenamefont {Wang}, \citenamefont {Sun},\ and\ \citenamefont
  {Miller}(1998)}]{Wang1998}%
  \BibitemOpen
  \bibfield  {author} {\bibinfo {author} {\bibfnamefont {H.}~\bibnamefont
  {Wang}}, \bibinfo {author} {\bibfnamefont {X.}~\bibnamefont {Sun}}, \ and\
  \bibinfo {author} {\bibfnamefont {W.~H.}\ \bibnamefont {Miller}},\ }\href
  {\doibase 10.1063/1.476447} {\bibfield  {journal} {\bibinfo  {journal} {The
  Journal of Chemical Physics}\ }\textbf {\bibinfo {volume} {108}},\ \bibinfo
  {pages} {9726} (\bibinfo {year} {1998})}\BibitemShut {NoStop}%
\bibitem [{\citenamefont {Cao}\ and\ \citenamefont
  {Voth}(1994{\natexlab{a}})}]{Cao1994-2}%
  \BibitemOpen
  \bibfield  {author} {\bibinfo {author} {\bibfnamefont {J.}~\bibnamefont
  {Cao}}\ and\ \bibinfo {author} {\bibfnamefont {G.~A.}\ \bibnamefont {Voth}},\
  }\href {\doibase 10.1063/1.467176} {\bibfield  {journal} {\bibinfo  {journal}
  {The Journal of Chemical Physics}\ }\textbf {\bibinfo {volume} {100}},\
  \bibinfo {pages} {5106} (\bibinfo {year} {1994}{\natexlab{a}})}\BibitemShut
  {NoStop}%
\bibitem [{\citenamefont {Cao}\ and\ \citenamefont
  {Voth}(1994{\natexlab{b}})}]{Cao1994-4}%
  \BibitemOpen
  \bibfield  {author} {\bibinfo {author} {\bibfnamefont {J.}~\bibnamefont
  {Cao}}\ and\ \bibinfo {author} {\bibfnamefont {G.~A.}\ \bibnamefont {Voth}},\
  }\href {\doibase 10.1063/1.468399} {\bibfield  {journal} {\bibinfo  {journal}
  {The Journal of Chemical Physics}\ }\textbf {\bibinfo {volume} {101}},\
  \bibinfo {pages} {6168} (\bibinfo {year} {1994}{\natexlab{b}})}\BibitemShut
  {NoStop}%
\bibitem [{\citenamefont {Jang}\ and\ \citenamefont {Voth}(1999)}]{Jang1999}%
  \BibitemOpen
  \bibfield  {author} {\bibinfo {author} {\bibfnamefont {S.}~\bibnamefont
  {Jang}}\ and\ \bibinfo {author} {\bibfnamefont {G.~A.}\ \bibnamefont
  {Voth}},\ }\href {\doibase 10.1063/1.479515} {\bibfield  {journal} {\bibinfo
  {journal} {The Journal of Chemical Physics}\ }\textbf {\bibinfo {volume}
  {111}},\ \bibinfo {pages} {2371} (\bibinfo {year} {1999})}\BibitemShut
  {NoStop}%
\bibitem [{\citenamefont {Craig}\ and\ \citenamefont
  {Manolopoulos}(2004)}]{Craig2004}%
  \BibitemOpen
  \bibfield  {author} {\bibinfo {author} {\bibfnamefont {I.~R.}\ \bibnamefont
  {Craig}}\ and\ \bibinfo {author} {\bibfnamefont {D.~E.}\ \bibnamefont
  {Manolopoulos}},\ }\href {\doibase 10.1063/1.1777575} {\bibfield  {journal}
  {\bibinfo  {journal} {The Journal of Chemical Physics}\ }\textbf {\bibinfo
  {volume} {121}},\ \bibinfo {pages} {3368} (\bibinfo {year}
  {2004})}\BibitemShut {NoStop}%
\bibitem [{\citenamefont {Habershon}\ \emph {et~al.}(2013)\citenamefont
  {Habershon}, \citenamefont {Manolopoulos}, \citenamefont {Markland},\ and\
  \citenamefont {Miller}}]{Habershon2013}%
  \BibitemOpen
  \bibfield  {author} {\bibinfo {author} {\bibfnamefont {S.}~\bibnamefont
  {Habershon}}, \bibinfo {author} {\bibfnamefont {D.~E.}\ \bibnamefont
  {Manolopoulos}}, \bibinfo {author} {\bibfnamefont {T.~E.}\ \bibnamefont
  {Markland}}, \ and\ \bibinfo {author} {\bibfnamefont {T.~F.}\ \bibnamefont
  {Miller}},\ }\href {\doibase 10.1146/annurev-physchem-040412-110122}
  {\bibfield  {journal} {\bibinfo  {journal} {Annual Review of Physical
  Chemistry}\ }\textbf {\bibinfo {volume} {64}},\ \bibinfo {pages} {387}
  (\bibinfo {year} {2013})}\BibitemShut {NoStop}%
\bibitem [{\citenamefont {Kubo}(1957)}]{Kubo1957}%
  \BibitemOpen
  \bibfield  {author} {\bibinfo {author} {\bibfnamefont {R.}~\bibnamefont
  {Kubo}},\ }\href {\doibase 10.1143/jpsj.12.570} {\bibfield  {journal}
  {\bibinfo  {journal} {Journal of the Physical Society of Japan}\ }\textbf
  {\bibinfo {volume} {12}},\ \bibinfo {pages} {570} (\bibinfo {year}
  {1957})}\BibitemShut {NoStop}%
\bibitem [{\citenamefont {Zwanzig}(1965)}]{Zwanzig_1965}%
  \BibitemOpen
  \bibfield  {author} {\bibinfo {author} {\bibfnamefont {R.}~\bibnamefont
  {Zwanzig}},\ }\href {\doibase 10.1146/annurev.pc.16.100165.000435} {\bibfield
   {journal} {\bibinfo  {journal} {Annual Review of Physical Chemistry}\
  }\textbf {\bibinfo {volume} {16}},\ \bibinfo {pages} {67} (\bibinfo {year}
  {1965})},\ \Eprint
  {http://arxiv.org/abs/https://doi.org/10.1146/annurev.pc.16.100165.000435}
  {https://doi.org/10.1146/annurev.pc.16.100165.000435} \BibitemShut {NoStop}%
\bibitem [{\citenamefont {Reichman}\ \emph {et~al.}(2000)\citenamefont
  {Reichman}, \citenamefont {Roy}, \citenamefont {Jang},\ and\ \citenamefont
  {Voth}}]{Reichman2000}%
  \BibitemOpen
  \bibfield  {author} {\bibinfo {author} {\bibfnamefont {D.~R.}\ \bibnamefont
  {Reichman}}, \bibinfo {author} {\bibfnamefont {P.-N.}\ \bibnamefont {Roy}},
  \bibinfo {author} {\bibfnamefont {S.}~\bibnamefont {Jang}}, \ and\ \bibinfo
  {author} {\bibfnamefont {G.~A.}\ \bibnamefont {Voth}},\ }\href {\doibase
  10.1063/1.481872} {\bibfield  {journal} {\bibinfo  {journal} {The Journal of
  Chemical Physics}\ }\textbf {\bibinfo {volume} {113}},\ \bibinfo {pages}
  {919} (\bibinfo {year} {2000})}\BibitemShut {NoStop}%
\bibitem [{\citenamefont {Jung}, \citenamefont {Videla},\ and\ \citenamefont
  {Batista}(2019)}]{Jung2019}%
  \BibitemOpen
  \bibfield  {author} {\bibinfo {author} {\bibfnamefont {K.~A.}\ \bibnamefont
  {Jung}}, \bibinfo {author} {\bibfnamefont {P.~E.}\ \bibnamefont {Videla}}, \
  and\ \bibinfo {author} {\bibfnamefont {V.~S.}\ \bibnamefont {Batista}},\
  }\href {\doibase 10.1063/1.5110427} {\bibfield  {journal} {\bibinfo
  {journal} {The Journal of Chemical Physics}\ }\textbf {\bibinfo {volume}
  {151}},\ \bibinfo {pages} {034108} (\bibinfo {year} {2019})}\BibitemShut
  {NoStop}%
\bibitem [{\citenamefont {Jung}, \citenamefont {Videla},\ and\ \citenamefont
  {Batista}(2018)}]{Jung2018}%
  \BibitemOpen
  \bibfield  {author} {\bibinfo {author} {\bibfnamefont {K.~A.}\ \bibnamefont
  {Jung}}, \bibinfo {author} {\bibfnamefont {P.~E.}\ \bibnamefont {Videla}}, \
  and\ \bibinfo {author} {\bibfnamefont {V.~S.}\ \bibnamefont {Batista}},\
  }\href {\doibase 10.1063/1.5036768} {\bibfield  {journal} {\bibinfo
  {journal} {The Journal of Chemical Physics}\ }\textbf {\bibinfo {volume}
  {148}},\ \bibinfo {pages} {244105} (\bibinfo {year} {2018})}\BibitemShut
  {NoStop}%
\bibitem [{\citenamefont {Tong}\ \emph {et~al.}(2020)\citenamefont {Tong},
  \citenamefont {Videla}, \citenamefont {Jung}, \citenamefont {Batista},\ and\
  \citenamefont {Sun}}]{Tong2020}%
  \BibitemOpen
  \bibfield  {author} {\bibinfo {author} {\bibfnamefont {Z.}~\bibnamefont
  {Tong}}, \bibinfo {author} {\bibfnamefont {P.~E.}\ \bibnamefont {Videla}},
  \bibinfo {author} {\bibfnamefont {K.~A.}\ \bibnamefont {Jung}}, \bibinfo
  {author} {\bibfnamefont {V.~S.}\ \bibnamefont {Batista}}, \ and\ \bibinfo
  {author} {\bibfnamefont {X.}~\bibnamefont {Sun}},\ }\href {\doibase
  10.1063/5.0015436} {\bibfield  {journal} {\bibinfo  {journal} {The Journal of
  Chemical Physics}\ }\textbf {\bibinfo {volume} {153}},\ \bibinfo {pages}
  {034117} (\bibinfo {year} {2020})}\BibitemShut {NoStop}%
\bibitem [{\citenamefont {Nitzan}(2006)}]{Nitzan_Book}%
  \BibitemOpen
  \bibfield  {author} {\bibinfo {author} {\bibfnamefont {A.}~\bibnamefont
  {Nitzan}},\ }\href@noop {} {\emph {\bibinfo {title} {Chemical Dynamics in
  Condensed Phases}}}\ (\bibinfo  {publisher} {Oxford University Press},\
  \bibinfo {year} {2006})\BibitemShut {NoStop}%
\bibitem [{\citenamefont {Jeon}\ and\ \citenamefont {Cho}(2010)}]{Jeon2010}%
  \BibitemOpen
  \bibfield  {author} {\bibinfo {author} {\bibfnamefont {J.}~\bibnamefont
  {Jeon}}\ and\ \bibinfo {author} {\bibfnamefont {M.}~\bibnamefont {Cho}},\
  }\href {\doibase 10.1088/1367-2630/12/6/065001} {\bibfield  {journal}
  {\bibinfo  {journal} {New Journal of Physics}\ }\textbf {\bibinfo {volume}
  {12}},\ \bibinfo {pages} {065001} (\bibinfo {year} {2010})}\BibitemShut
  {NoStop}%
\bibitem [{\citenamefont {Jeon}\ and\ \citenamefont {Cho}(2014)}]{Jeon2014}%
  \BibitemOpen
  \bibfield  {author} {\bibinfo {author} {\bibfnamefont {J.}~\bibnamefont
  {Jeon}}\ and\ \bibinfo {author} {\bibfnamefont {M.}~\bibnamefont {Cho}},\
  }\href {\doibase 10.1021/jp501182d} {\bibfield  {journal} {\bibinfo
  {journal} {The Journal of Physical Chemistry B}\ }\textbf {\bibinfo {volume}
  {118}},\ \bibinfo {pages} {8148} (\bibinfo {year} {2014})}\BibitemShut
  {NoStop}%
\bibitem [{\citenamefont {P\'{e}rez}, \citenamefont {Tuckerman},\ and\
  \citenamefont {M\"{u}ser}(2009)}]{Perez2009}%
  \BibitemOpen
  \bibfield  {author} {\bibinfo {author} {\bibfnamefont {A.}~\bibnamefont
  {P\'{e}rez}}, \bibinfo {author} {\bibfnamefont {M.~E.}\ \bibnamefont
  {Tuckerman}}, \ and\ \bibinfo {author} {\bibfnamefont {M.~H.}\ \bibnamefont
  {M\"{u}ser}},\ }\href {\doibase 10.1063/1.3126950} {\bibfield  {journal}
  {\bibinfo  {journal} {The Journal of Chemical Physics}\ }\textbf {\bibinfo
  {volume} {130}},\ \bibinfo {pages} {184105} (\bibinfo {year}
  {2009})}\BibitemShut {NoStop}%
\bibitem [{\citenamefont {Habershon}, \citenamefont {Markland},\ and\
  \citenamefont {Manolopoulos}(2009)}]{Habershon2009}%
  \BibitemOpen
  \bibfield  {author} {\bibinfo {author} {\bibfnamefont {S.}~\bibnamefont
  {Habershon}}, \bibinfo {author} {\bibfnamefont {T.~E.}\ \bibnamefont
  {Markland}}, \ and\ \bibinfo {author} {\bibfnamefont {D.~E.}\ \bibnamefont
  {Manolopoulos}},\ }\href {\doibase 10.1063/1.3167790} {\bibfield  {journal}
  {\bibinfo  {journal} {The Journal of Chemical Physics}\ }\textbf {\bibinfo
  {volume} {131}},\ \bibinfo {pages} {024501} (\bibinfo {year} {2009})},\
  \Eprint {http://arxiv.org/abs/https://doi.org/10.1063/1.3167790}
  {https://doi.org/10.1063/1.3167790} \BibitemShut {NoStop}%
\bibitem [{\citenamefont {Rossi}, \citenamefont {Ceriotti},\ and\ \citenamefont
  {Manolopoulos}(2014)}]{Rossi2014}%
  \BibitemOpen
  \bibfield  {author} {\bibinfo {author} {\bibfnamefont {M.}~\bibnamefont
  {Rossi}}, \bibinfo {author} {\bibfnamefont {M.}~\bibnamefont {Ceriotti}}, \
  and\ \bibinfo {author} {\bibfnamefont {D.~E.}\ \bibnamefont {Manolopoulos}},\
  }\href {\doibase 10.1063/1.4883861} {\bibfield  {journal} {\bibinfo
  {journal} {The Journal of Chemical Physics}\ }\textbf {\bibinfo {volume}
  {140}},\ \bibinfo {pages} {234116} (\bibinfo {year} {2014})}\BibitemShut
  {NoStop}%
\bibitem [{\citenamefont {Hele}(2016)}]{Hele2016}%
  \BibitemOpen
  \bibfield  {author} {\bibinfo {author} {\bibfnamefont {T.~J.~H.}\
  \bibnamefont {Hele}},\ }\href {\doibase 10.1080/00268976.2015.1136003}
  {\bibfield  {journal} {\bibinfo  {journal} {Molecular Physics}\ }\textbf
  {\bibinfo {volume} {114}},\ \bibinfo {pages} {1461} (\bibinfo {year}
  {2016})}\BibitemShut {NoStop}%
\bibitem [{\citenamefont {Marston}\ and\ \citenamefont
  {Balint-Kurti}(1989)}]{Marston1989}%
  \BibitemOpen
  \bibfield  {author} {\bibinfo {author} {\bibfnamefont {C.~C.}\ \bibnamefont
  {Marston}}\ and\ \bibinfo {author} {\bibfnamefont {G.~G.}\ \bibnamefont
  {Balint-Kurti}},\ }\href {\doibase 10.1063/1.456888} {\bibfield  {journal}
  {\bibinfo  {journal} {The Journal of Chemical Physics}\ }\textbf {\bibinfo
  {volume} {91}},\ \bibinfo {pages} {3571} (\bibinfo {year}
  {1989})}\BibitemShut {NoStop}%
\bibitem [{\citenamefont {Hele}\ \emph {et~al.}(2015)\citenamefont {Hele},
  \citenamefont {Willatt}, \citenamefont {Muolo},\ and\ \citenamefont
  {Althorpe}}]{Hele2015}%
  \BibitemOpen
  \bibfield  {author} {\bibinfo {author} {\bibfnamefont {T.~J.~H.}\
  \bibnamefont {Hele}}, \bibinfo {author} {\bibfnamefont {M.~J.}\ \bibnamefont
  {Willatt}}, \bibinfo {author} {\bibfnamefont {A.}~\bibnamefont {Muolo}}, \
  and\ \bibinfo {author} {\bibfnamefont {S.~C.}\ \bibnamefont {Althorpe}},\
  }\href {\doibase 10.1063/1.4921234} {\bibfield  {journal} {\bibinfo
  {journal} {The Journal of Chemical Physics}\ }\textbf {\bibinfo {volume}
  {142}},\ \bibinfo {pages} {191101} (\bibinfo {year} {2015})}\BibitemShut
  {NoStop}%
\bibitem [{\citenamefont {Jung}, \citenamefont {Videla},\ and\ \citenamefont
  {Batista}(2020)}]{Jung2020}%
  \BibitemOpen
  \bibfield  {author} {\bibinfo {author} {\bibfnamefont {K.~A.}\ \bibnamefont
  {Jung}}, \bibinfo {author} {\bibfnamefont {P.~E.}\ \bibnamefont {Videla}}, \
  and\ \bibinfo {author} {\bibfnamefont {V.~S.}\ \bibnamefont {Batista}},\
  }\href {\doibase 10.1063/5.0021843} {\bibfield  {journal} {\bibinfo
  {journal} {The Journal of Chemical Physics}\ }\textbf {\bibinfo {volume}
  {153}},\ \bibinfo {pages} {124112} (\bibinfo {year} {2020})}\BibitemShut
  {NoStop}%
\bibitem [{\citenamefont {Noda}(2009)}]{Noda2009}%
  \BibitemOpen
  \bibfield  {author} {\bibinfo {author} {\bibfnamefont {I.}~\bibnamefont
  {Noda}},\ }in\ \href {\doibase 10.1016/b978-0-444-53175-9.00013-1} {\emph
  {\bibinfo {booktitle} {Frontiers of Molecular Spectroscopy}}}\ (\bibinfo
  {publisher} {Elsevier},\ \bibinfo {year} {2009})\ pp.\ \bibinfo {pages}
  {367--381}\BibitemShut {NoStop}%
\bibitem [{\citenamefont {Morawietz}\ \emph {et~al.}(2018)\citenamefont
  {Morawietz}, \citenamefont {Marsalek}, \citenamefont {Pattenaude},
  \citenamefont {Streacker}, \citenamefont {Ben-Amotz},\ and\ \citenamefont
  {Markland}}]{Morawietz2018}%
  \BibitemOpen
  \bibfield  {author} {\bibinfo {author} {\bibfnamefont {T.}~\bibnamefont
  {Morawietz}}, \bibinfo {author} {\bibfnamefont {O.}~\bibnamefont {Marsalek}},
  \bibinfo {author} {\bibfnamefont {S.~R.}\ \bibnamefont {Pattenaude}},
  \bibinfo {author} {\bibfnamefont {L.~M.}\ \bibnamefont {Streacker}}, \bibinfo
  {author} {\bibfnamefont {D.}~\bibnamefont {Ben-Amotz}}, \ and\ \bibinfo
  {author} {\bibfnamefont {T.~E.}\ \bibnamefont {Markland}},\ }\href {\doibase
  10.1021/acs.jpclett.8b00133} {\bibfield  {journal} {\bibinfo  {journal} {The
  Journal of Physical Chemistry Letters}\ }\textbf {\bibinfo {volume} {9}},\
  \bibinfo {pages} {851} (\bibinfo {year} {2018})}\BibitemShut {NoStop}%
\bibitem [{\citenamefont {Markland}\ and\ \citenamefont
  {Ceriotti}(2018)}]{Markland2018}%
  \BibitemOpen
  \bibfield  {author} {\bibinfo {author} {\bibfnamefont {T.~E.}\ \bibnamefont
  {Markland}}\ and\ \bibinfo {author} {\bibfnamefont {M.}~\bibnamefont
  {Ceriotti}},\ }\href {\doibase 10.1038/s41570-017-0109} {\bibfield  {journal}
  {\bibinfo  {journal} {Nature Reviews Chemistry}\ }\textbf {\bibinfo {volume}
  {2}} (\bibinfo {year} {2018}),\ 10.1038/s41570-017-0109}\BibitemShut
  {NoStop}%
\bibitem [{\citenamefont {Th\"{a}mer}\ \emph {et~al.}(2015)\citenamefont
  {Th\"{a}mer}, \citenamefont {Marco}, \citenamefont {Ramasesha}, \citenamefont
  {Mandal},\ and\ \citenamefont {Tokmakoff}}]{Thmer2015}%
  \BibitemOpen
  \bibfield  {author} {\bibinfo {author} {\bibfnamefont {M.}~\bibnamefont
  {Th\"{a}mer}}, \bibinfo {author} {\bibfnamefont {L.~D.}\ \bibnamefont
  {Marco}}, \bibinfo {author} {\bibfnamefont {K.}~\bibnamefont {Ramasesha}},
  \bibinfo {author} {\bibfnamefont {A.}~\bibnamefont {Mandal}}, \ and\ \bibinfo
  {author} {\bibfnamefont {A.}~\bibnamefont {Tokmakoff}},\ }\href {\doibase
  10.1126/science.aab3908} {\bibfield  {journal} {\bibinfo  {journal}
  {Science}\ }\textbf {\bibinfo {volume} {350}},\ \bibinfo {pages} {78}
  (\bibinfo {year} {2015})}\BibitemShut {NoStop}%
\bibitem [{\citenamefont {Yuan}\ and\ \citenamefont {Fayer}(2019)}]{Yuan2019}%
  \BibitemOpen
  \bibfield  {author} {\bibinfo {author} {\bibfnamefont {R.}~\bibnamefont
  {Yuan}}\ and\ \bibinfo {author} {\bibfnamefont {M.~D.}\ \bibnamefont
  {Fayer}},\ }\href {\doibase 10.1021/acs.jpcb.9b06038} {\bibfield  {journal}
  {\bibinfo  {journal} {The Journal of Physical Chemistry B}\ }\textbf
  {\bibinfo {volume} {123}},\ \bibinfo {pages} {7628} (\bibinfo {year}
  {2019})}\BibitemShut {NoStop}%
\end{thebibliography}%
\end{document}


\title{Supplementary Material: 2D spectroscopies from condensed phase quantum dynamics: Accessing third-order response properties from equilibrium multi-time correlation functions}

\author{Kenneth A. Jung}
\author{Thomas E. Markland}
\email{tmarkland@stanford.edu}
\affiliation{Department of Chemistry, Stanford University, Stanford, California, 94305, USA}

\date{\today}

\maketitle

\section{Limits of the  Kubo frequency factor: $f(\omega_1,\omega_2,\omega_3)$}
The general three-time Kubo frequency factor is given as
\begin{equation}
f(\omega_1,\omega_2,\omega_3) = \frac{1}{\beta^3}\int^{\beta}_0 d\lambda \int^{\lambda}_0 d\mu \int^{\mu}_0 d\nu \: e^{\lambda\hbar a} e^{\mu\hbar b} e^{\nu\hbar c},
\label{eq:F_factor}
\end{equation}
where particular choices of $a$, $b$, and $c$ gives a particular conversion factor between standard and Kubo transformed three-time correlation functions. For example, if $a=\omega_2$, $b=\omega_3$, and $c=\omega_1$ we get the conversion factor between $S$ and $K$ in the main text. Evaluating the integrals in Eq.~(\ref{eq:F_factor}) we obtain
\begin{equation}
f(\omega_1,\omega_2,\omega_3) = \frac{1}{\beta^3\hbar^3}\left[ \frac{e^{\beta\hbar(a+b+c)}-1}{c(b+c)(a+b+c)} 
+ \frac{e^{\beta\hbar a}-1}{ab(b+c)} 
- \frac{e^{\beta\hbar(a+b)}-1}{bc(a+b)}
\right].
\label{eq:F_factor_red}
\end{equation}
Numerically applying Eq.~(\ref{eq:F_factor_red}) as written can be difficult since there appears to be singularities when either $a=0$, $b=0$, $c=0$, $a=-b$, $b=-c$, or $a=-(b+c)$ and hence care must be taken in applying these limits. When these limits are appropriately taken one obtains:
\begin{equation}
\lim_{a\to0}f = \frac{b^2(e^{\beta\hbar(b+c)}-e^{\beta\hbar b} + \beta\hbar c) - c^2(e^{\beta\hbar b}-1 ) +bc(2+\beta\hbar c -2e^{\beta\hbar b})}{\beta^3\hbar^3b^2c(b+c)^2},
\end{equation}
\begin{equation}
\lim_{b\to0}f = \frac{a^2e^{\beta\hbar a}(e^{\beta\hbar c}-1)-\beta\hbar a^2ce^{\beta\hbar a} +c^2(e^{\beta\hbar a}(1-\beta\hbar a)-1)}{\beta^3\hbar^3a^2c^2(a+c)},
\end{equation}
\begin{equation}
\lim_{c\to0}f = \frac{(a+b)^2(e^{\beta\hbar a}-1) + \beta\hbar ab(a+b)e^{\beta\hbar(a+b)}-a(a+2b)(e^{\beta\hbar(a+b)}-1)}{\beta^3\hbar^3ab^2(a+b)^2},
\end{equation}
\begin{equation}
\lim_{a\to-b}f = \frac{c^2(1-e^{-\beta\hbar b})-\beta\hbar bc^2+b^2(e^{\beta\hbar c} - \beta\hbar c -1)}{\beta^3\hbar^3b^2c^2(b+c)},
\end{equation}
\begin{equation}
\lim_{b\to-c}f = \frac{-a^2(e^{\beta\hbar a}-e^{\beta\hbar(a-c)}) +\beta\hbar a^2ce^{\beta\hbar a}+c^2(e^{\beta\hbar a}(1-\beta\hbar a)-1)}{\beta^3\hbar^3a^2c^2(a-c)},
\end{equation}
\begin{equation}
\lim_{a\to-(b+c)}f = \frac{-(b+c)(ce^{-\beta\hbar(b+c)}+b-(b+c)e^{-\beta\hbar c}) + bce^{-\beta\hbar(b+c)} + \beta\hbar(b+c) -1}{\beta^3\hbar^3bc^2(b+c)^2},
\end{equation}
\begin{equation}
\lim_{a\to0,b\to0}f = -\frac{2-2e^{\beta\hbar c}+2c\beta\hbar + \beta^2\hbar^2c^2}{2\beta^3\hbar^3c^3},
\end{equation}
\begin{equation}
\lim_{a\to0,c\to0}f = \frac{2+\beta\hbar b + e^{\beta\hbar b}(\beta\hbar b -2)}{\beta^3\hbar^3b^3},
\end{equation}
\begin{equation}
\lim_{b\to0,c\to0}f = \frac{e^{\beta\hbar a}(2-2\beta\hbar a + \beta^2\hbar^2a^2)-2}{2\beta^3\hbar^3a^3},
\end{equation}
\begin{equation}
\lim_{a\to0,b\to-c}f = \frac{2-2e^{-\beta\hbar c}-2\beta\hbar c + \beta^2\hbar^2c^2}{2\beta^3\hbar^3c^3},
\end{equation}
\begin{equation}
\lim_{a\to-b,c\to0}f = \frac{2-2e^{-\beta\hbar b}-2\beta\hbar b + \beta^2\hbar^2b^2}{2\beta^3\hbar^3b^3},
\end{equation}
\begin{equation}
\lim_{a\to-b,b\to c}f = \frac{\frac{c}{b}+ \frac{2c}{c-b} - \frac{c}{2c-b}-\frac{ce^{-\beta\hbar b}}{b} +e^{\beta\hbar(c-b)}\left( \frac{ce^{\beta\hbar c}}{2c-b} - \frac{2c}{c-b} \right) }{2\beta^3\hbar^3c^3},
\end{equation}
\begin{equation}
\lim_{a\to-b,b\to-c}f = \frac{2+\beta\hbar c +e^{\beta\hbar c}(\beta\hbar c -2)}{\beta^3\hbar^3c^3},
\end{equation}
\begin{equation}
\lim_{a\to-(b+c),b\to0}f = \frac{e^{\beta\hbar c}(2+\beta\hbar c +e^{\beta\hbar c}(\beta\hbar c -2)) }{\beta^3\hbar^3c^3},
\end{equation}
\begin{equation}
\lim_{a\to0,b\to0,c\to0}f = \frac{1}{6}.
\label{eq:Class_limit}
\end{equation}
Equation (\ref{eq:Class_limit}) is of particular importance since it yields the classical limit of the Kubo frequency factor.

\section{Harmonic Limit of $K$ and derivation of the $b(\omega_1,\omega_2,\omega_3)$ factor}

To provide a connection between the symmetrized Kubo transformed correlation function (SKTCF) and $\tilde{K}(\omega_1,\omega_2,\omega_3)$ we consider their connection in the limit of a harmonic potential ($V(\hat{q}) = \frac{m\Omega^2}{2}\hat{q}^2$) with linear operators since this is the simplest model that gives a nonzero three-time correlation function.  

The harmonic limit of the SKTCF for linear operators is,  
\begin{equation}
 K^{\textrm{sym}}(t_1,t_2,t_3) = \frac{1}{(\beta m \Omega^2)^2}\left\{ \cos[\Omega(t_1-t_2-t_3)] + \cos[\Omega(t_1+t_2-t_3)] + \cos[\Omega(t_1-t_2+t_3)] 
\right\}.
\label{eq:part0}
\end{equation}
The harmonic limit of $K(t_1,t_2,t_3)$ is a sum of six terms \begin{equation}
K = K_{\uparrow\downarrow\uparrow\downarrow} +K_{\downarrow\uparrow\downarrow\uparrow}
+ 
K_{\uparrow\uparrow\downarrow\downarrow} + K_{\downarrow\downarrow\uparrow\uparrow} + K_{\uparrow\downarrow\downarrow\uparrow} + K_{\downarrow\uparrow\uparrow\downarrow}.
\label{eq:ladderexpression}
\end{equation}
The arrows indicate a contributing term from the combinations of the raising or lowering that stem from the four $\hat{q}=\sqrt{\frac{\hbar}{2m\Omega}}(\hat{a}+\hat{a}^{\dagger})$ operators involved in the correlation function. Any other combination than those appearing in Eq.~(\ref{eq:ladderexpression}) vanish. Each of the terms in Eq.~(\ref{eq:ladderexpression}) can be evaluated to give:
\begin{eqnarray}
K_{\uparrow\downarrow\uparrow\downarrow} &=& \frac{\coth(\beta\hbar\Omega/2)[\beta\hbar\Omega\coth(\beta\hbar\Omega/2)-2]}{4\beta^3\Omega^5\hbar}e^{-i\Omega(t_1-t_2+t_3)/\hbar},
\label{eq:start1}
\end{eqnarray}
\begin{eqnarray}
K_{\downarrow\uparrow\downarrow\uparrow} &=& \frac{\coth(\beta\hbar\Omega/2)[\beta\hbar\Omega\coth(\beta\hbar\Omega/2)-2]}{4\beta^3\Omega^5\hbar}e^{i\Omega(t_1-t_2+t_3)/\hbar},
\end{eqnarray}
\begin{eqnarray}
K_{\uparrow\uparrow\downarrow\downarrow} &=& \frac{\textrm{csch}(\beta\hbar\Omega/2)^2[\sinh(\beta\hbar\Omega/2)-\beta\hbar\Omega]}{8\beta^3\Omega^5\hbar}e^{-i\Omega(t_1+t_2-t_3)/\hbar},
\end{eqnarray}
\begin{eqnarray}
K_{\downarrow\downarrow\uparrow\uparrow} &=& \frac{\textrm{csch}(\beta\hbar\Omega/2)^2[\sinh(\beta\hbar\Omega/2)-\beta\hbar\Omega]}{8\beta^3\Omega^5\hbar}e^{i\Omega(t_1+t_2-t_3)/\hbar},
\end{eqnarray}
\begin{eqnarray}
K_{\uparrow\downarrow\downarrow\uparrow} &=& \frac{\textrm{csch}(\beta\hbar\Omega/2)^2[\sinh(\beta\hbar\Omega/2)-\beta\hbar\Omega]}{8\beta^3\Omega^5\hbar}e^{-i\Omega(t_1-t_2-t_3)/\hbar},
\end{eqnarray}
\begin{equation}
K_{\downarrow\uparrow\uparrow\downarrow} = \frac{\textrm{csch}(\beta\hbar\Omega/2)^2[\sinh(\beta\hbar\Omega/2)-\beta\hbar\Omega]}{8\beta^3\Omega^5\hbar}e^{i\Omega(t_1-t_2-t_3)/\hbar}.
\label{eq:end1}
\end{equation}
From Eqs.~(\ref{eq:start1})-(\ref{eq:end1}) it is immediately clear that imaginary part of $K(t_1,t_2,t_3)$ is zero. The real part is given by 
\begin{eqnarray}
\textrm{Re}[K(t_1,t_2,t_3)] &=& \frac{\textrm{csch}(\beta\hbar\Omega/2)^2[\sinh(\beta\hbar\Omega/2)-\beta\hbar\Omega]}{4\beta^3\Omega^5\hbar}\left\{ \cos[\Omega(t_1-t_2-t_3)] + \cos[\Omega(t_1+t_2-t_3)] \right\} +\nonumber \\
& & \frac{\coth(\beta\hbar\Omega/2)[\beta\hbar\Omega\coth(\beta\hbar\Omega/2)-2]}{4\beta^3\Omega^5\hbar} \cos[\Omega(t_1-t_2+t_3)].
\label{eq:part1}
\end{eqnarray}
To determine the $b(\omega_1,\omega_2,\omega_3)$ we follow the procedure taken in past works\cite{DeVane2003,DeVane2004,Jung2018} by taking the Fourier transforms of Eq.~(\ref{eq:part1}) and Eq.~(\ref{eq:part0}) then looking at the ratio and comparing terms. The simplest factor that gets two of the three terms perfectly and very nearly captures the other term is given by
\begin{equation}
b(\omega_1,\omega_2,\omega_3) = \frac{\sinh(\beta\hbar\bar{\omega}) - \beta\hbar\bar{\omega}}{\beta\hbar\bar{\omega}[\cosh(\beta\hbar\bar{\omega})-1]},
\end{equation}
which is the conversion factor given in the main text. It is straightforward to verify that $b(\omega_1,\omega_2,\omega_3)$ is an even function as it must be since both $\textrm{Re}[\tilde{K}(\omega_1,\omega_2,\omega_3)]$ and $\tilde{K}^{\textrm{sym}}(\omega_1,\omega_2,\omega_3)$ are even.

\section{Additional Figures}

\begin{figure}[b]
    \begin{center}
        \includegraphics[width=\textwidth]{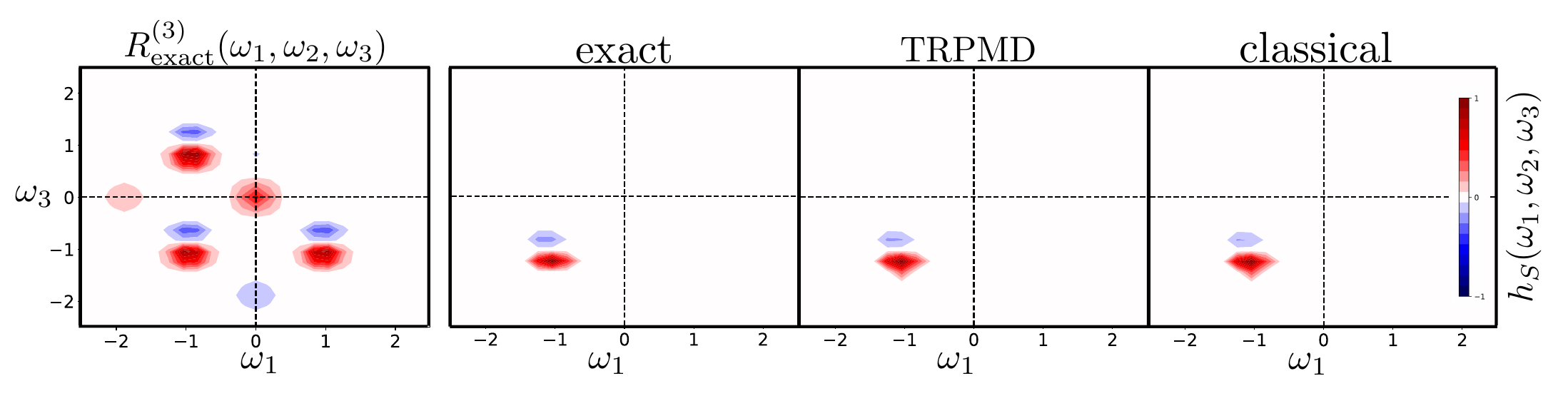}
    \end{center}
    \vspace{-5mm}
    \caption{Third-order order response for the MAP model at $\omega_2=1$ and $\beta=8$. The $h_S(\omega_1,\omega_2,\omega_3)$ correction factor derived in Appendix D is used here. The columns correspond to the levels of dynamics used to which the correction factor is applied to yield the third-order response: exact quantum dynamics (first column), TRPMD (second column), and classical dynamics (third column).}
\end{figure}

\begin{figure}[h!]
    \begin{center}
        \includegraphics[width=.8\textwidth]{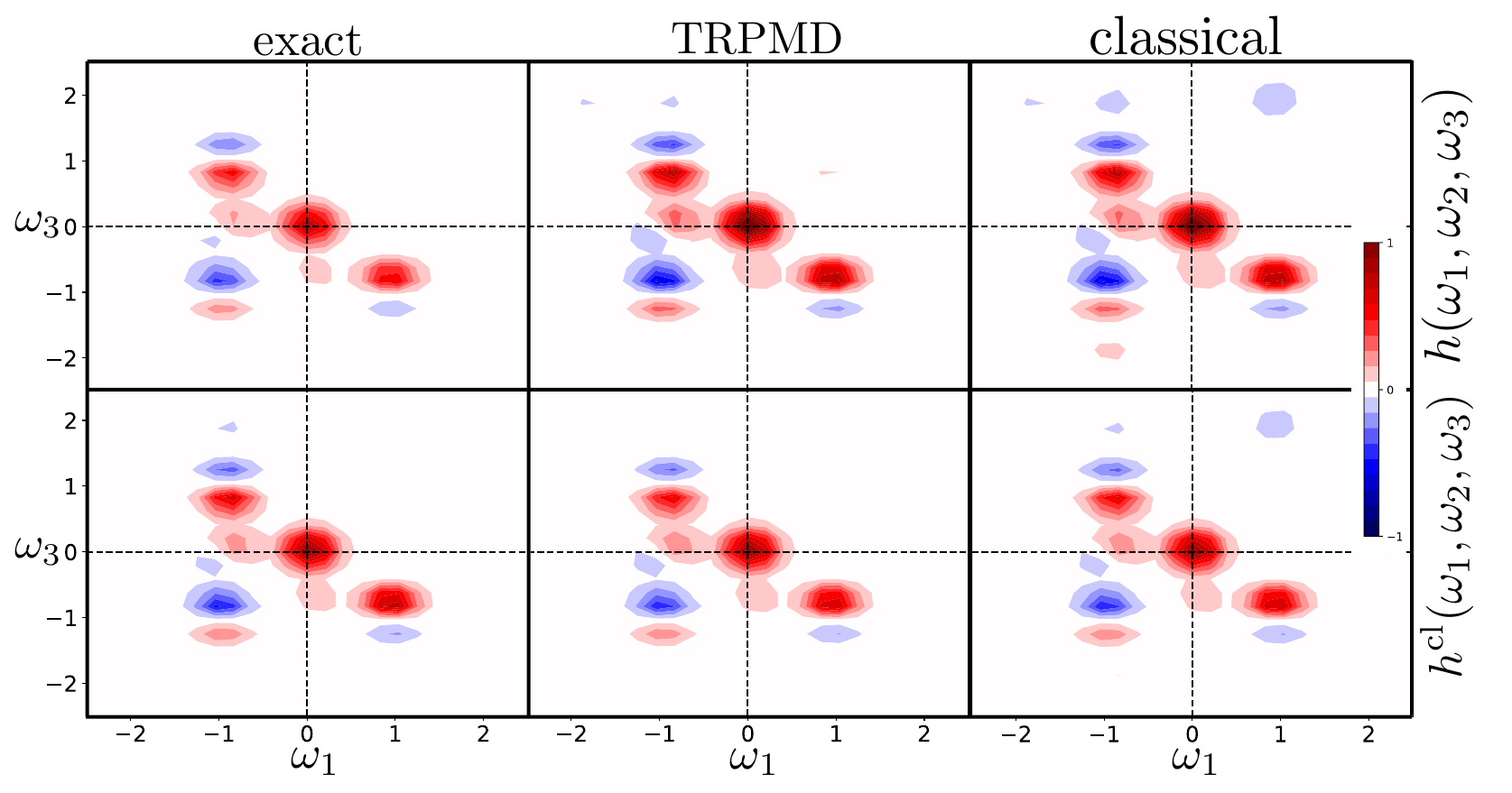}
    \end{center}
    \vspace{-5mm}
   \caption{Third-order response obtained for the MAP model at $\beta=1$ and $\omega_2=1$ using the full correction factor (top row) and its classical limit (bottom row). The columns show the effect of three levels of dynamics used to compute the SKTCF to which the correction factor is applied to yield the third-order response: exact quantum dynamics (first column), TRPMD (second column), and classical dynamics (third column).}
\end{figure}

\begin{figure}[h!]
    \begin{center}
        \includegraphics[width=\textwidth]{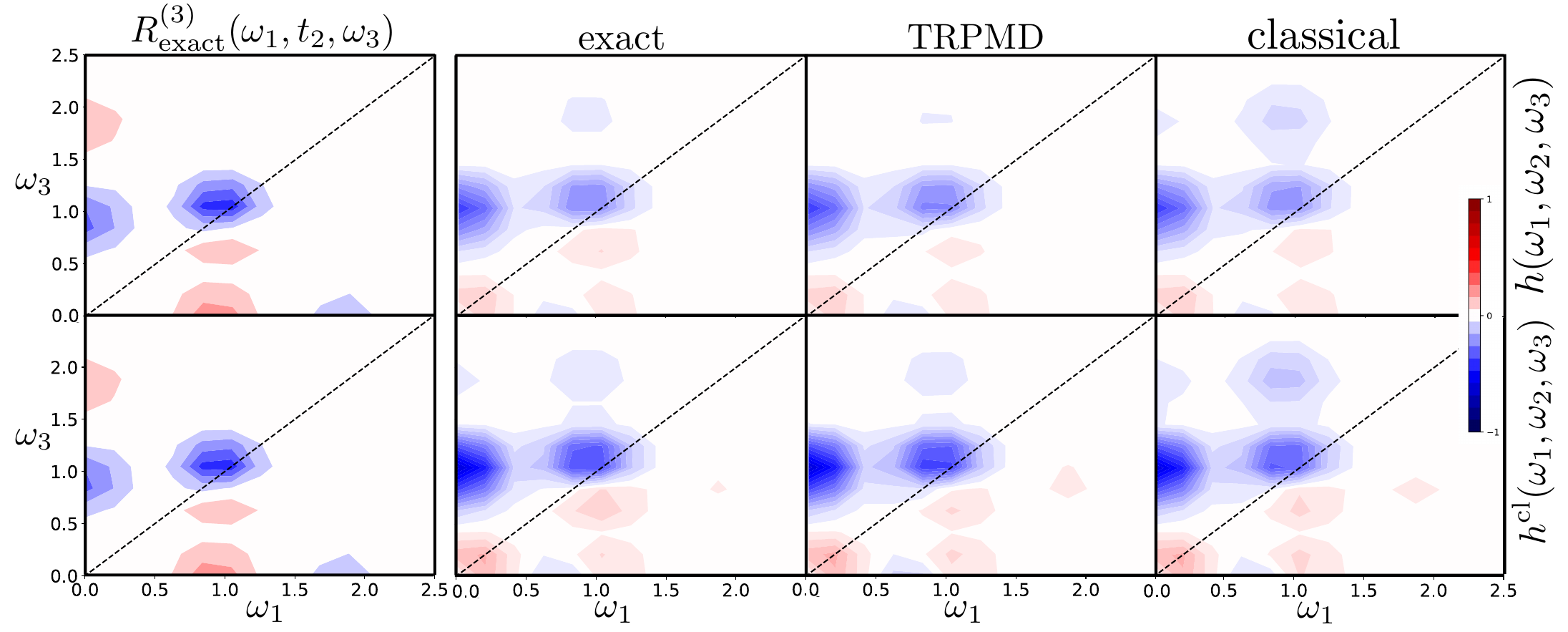}
    \end{center}
    \vspace{-5mm}
    \caption{Third-order order response for the MAP model with the $\omega_2$ axis transformed to the time domain, $R^{(3)}(\omega_1,t_2,\omega_3)$ at $t_2=0$ and $\beta=1$. The top row uses the full correction factor while the bottom row uses its classical limit. The columns correspond to the levels of dynamics used to which the correction factor is applied to yield the third-order response: exact quantum dynamics (first column), TRPMD (second column), and classical dynamics (third column).}
\end{figure}

\begin{figure}[h!]
    \begin{center}
        \includegraphics[width=.8\textwidth]{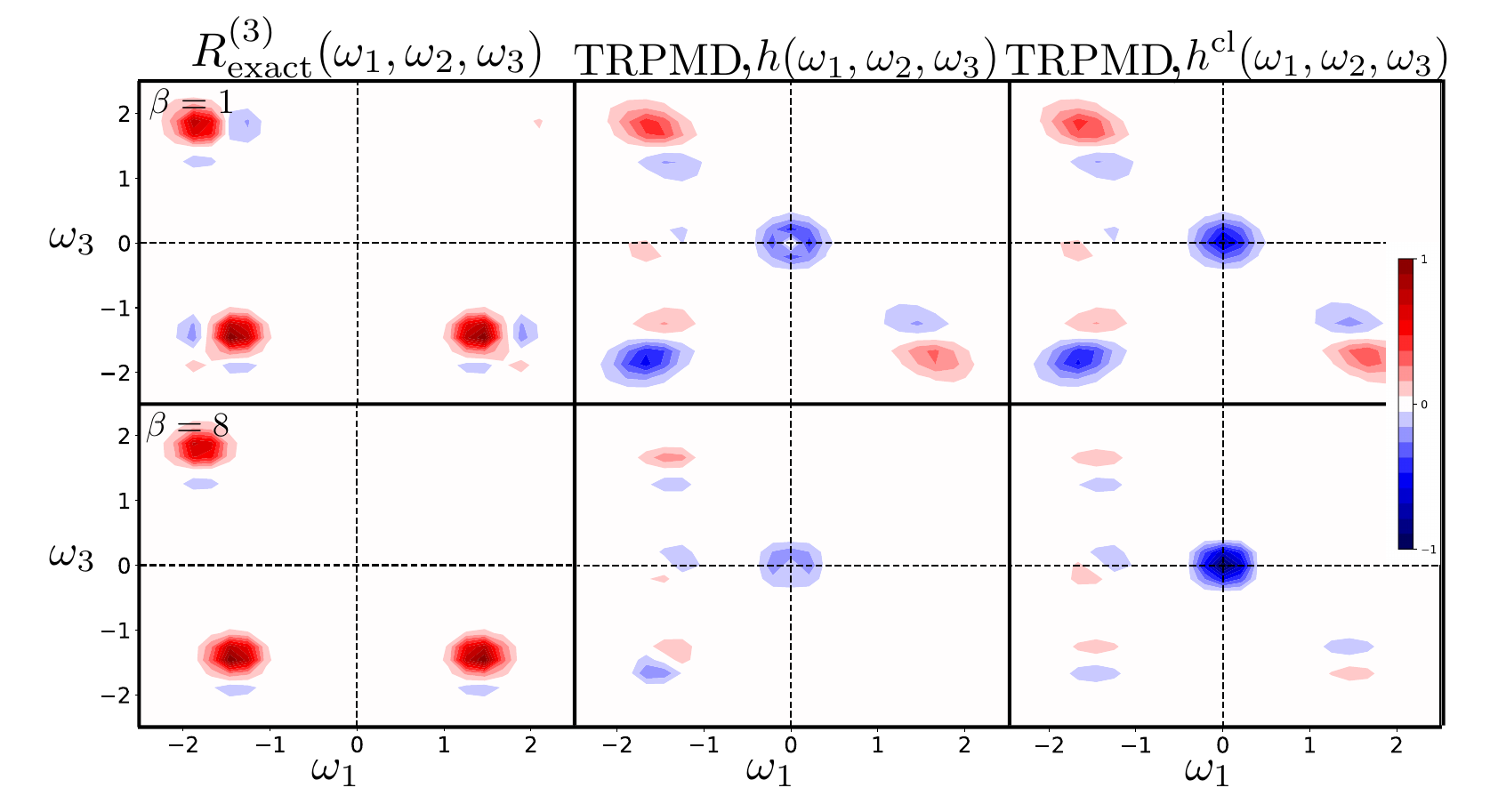}
    \end{center}
    \vspace{-5mm}
    \caption{Third-order response for the more strongly anharmonic OHP model at $\beta=1$ (top row) and $\beta=8$ (bottom row) at $w_2=1$. The left column contains the exact third-order response while the right column shows the third-order response obtained using the TRPMD approximated SKTCF in conjunction with the full correction factor.}
\end{figure}

\begin{figure}[h!]
    \begin{center}
        \includegraphics[width=.8\textwidth]{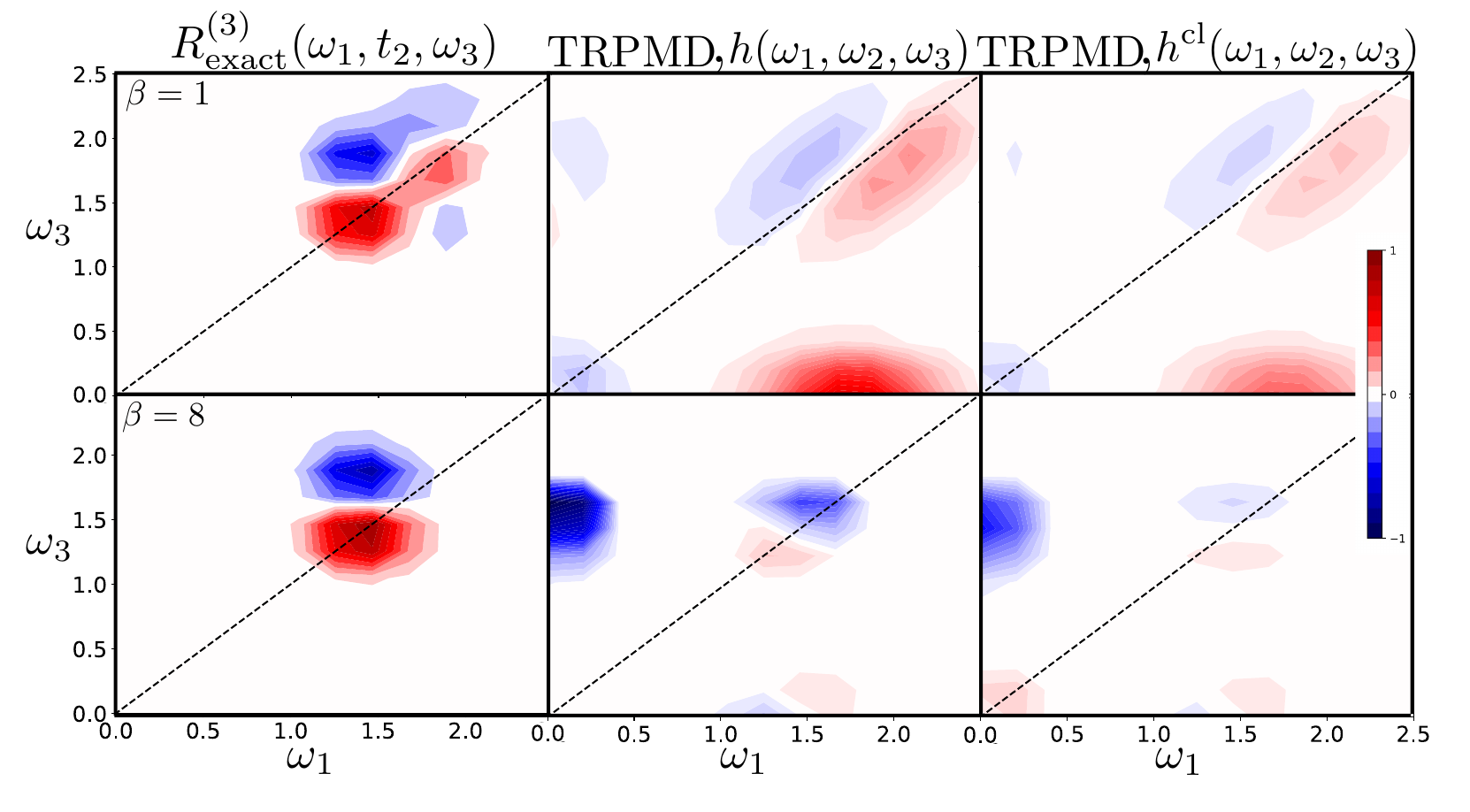}
    \end{center}
    \vspace{-5mm}
    \caption{Third-order response for the more strongly anharmonic OHP model at $\beta=1$ (top row) and $\beta=8$ (bottom row) at $t_2=0$. The left column contains the exact third-order response while the middle column shows the third-order response obtained using the TRPMD approximated SKTCF in conjunction with the full correction factor and the right column is the result of using the classical limit of the correction factor.}
\end{figure}

\clearpage

\bibliography{QCF}